\documentclass[journal,twocolumn]{IEEEtranTCOM}
\usepackage{array,cite}
\usepackage{graphicx,epsfig,subfigure}
\usepackage{amsthm}
\usepackage{amsmath}
\usepackage{mdwmath}
\usepackage{array}
\usepackage{subeqnarray}
\usepackage{stfloats}
\usepackage{algorithm,algorithmic}
\usepackage{xcolor}
\usepackage{amssymb}
\usepackage{upgreek}
\usepackage{threeparttable}

\newtheorem{theorem}{Theorem}
\newtheorem{fact}{Fact}
\newtheorem{lemma}{Lemma}

\begin{document}

\title{
Differential Modulation Exploiting the Spatial-Temporal Correlation of Wireless Channels With Moving Antenna Array}

\author{\IEEEauthorblockN{Zhaoyang~Zhang, \textit{Member, IEEE}, Chunxu~Jiao, Caijun~Zhong, \textit{Senior Member, IEEE},\\
Huazi~Zhang, \textit{Member, IEEE}, and Yu~Zhang, \textit{Member, IEEE}}
\thanks{
This work was supported in part by National Key Basic Research Program of China (No. 2012CB316104), National Hi-Tech R\&D Program of China (No.2014AA01A702), National Natural Science Foundation of China (Nos. 61371094, 61401391, 61401388), Zhejiang Provincial Natural Science Foundation of China (Nos. LR12F01002, LR15F010001), Zhejiang Science and Technology Department Public Project (2014C31051), and China Postdoctoral Science Foundation Funded Project (No. 2014M551736).

Zhaoyang~Zhang, Chunxu~Jiao, Caijun~Zhong and Huazi~Zhang are with the College of Information Science and Electronic Engineering, Zhejiang University, Hangzhou 310027, China. E-mails: \{ning\_ming, jiaocx1990, caijunzhong, hzhang17\}@zju.edu.cn.

Yu~Zhang is with the College of Information Engineering, Zhejiang University of Technology, Hangzhou 310014, China, and Provincial Key Laboratory of Information Networks, Zhejiang, China. E-mail: zhangyu\_wing@hotmail.com.}
}
\maketitle

\begin{abstract}
Provisioning reliable wireless services for railway passengers is becoming an increasingly critical problem to be addressed with the fast development of high speed trains (HST). In this paper, exploiting the linear mobility inherent to the HST communication scenario, we discover a new type of spatial-temporal correlation between the base station and moving antenna array on the roof top of the train. Capitalizing on the new spatial-temporal correlation structure and properties, an improved differential space-time modulation (DSTM) scheme is proposed. Analytical expressions are obtained for the pairwise error probability of the system. It is demonstrated that, the proposed approach achieves superior error performance compared with the conventional DSTM scheme. In addition, an adaptive method which dynamically adjusts the transmission block length is proposed to further enhance the system performance. Numerical results are provided to verify the performance of the proposed schemes.
\end{abstract}

%
\section{Introduction}
With the fast development of the high speed train (HST) networks around the world, particularly in China, Japan and Europe, provisioning high speed wireless connections to railway passengers has become an increasingly pressing issue to be tackled. However, moving at a speed of 300-500km/h poses significant difficulties for the design of efficient and reliable wireless systems due to two major obstacles, i.e., extremely frequent handovers and fast variation of the channel conditions \cite{calle2013long}. One of the promising solutions to address these challenges is to exploit the spatial diversity by implementing an antenna array on the top of the train \cite{luo2013efficient,ghazal2012non}.

The potential benefits offered by multiple antenna technology depend heavily on the availability of channel state information (CSI) at the communication nodes. Nevertheless, in the high mobility scenario, obtaining accurate CSI is a difficult task, hence a natural choice is to adopt the differential space-time modulation (DSTM) scheme, which eliminates the requirement for continuous tracking of the CSI \cite{hughes2000differential}. Although DSTM for MIMO systems has been well studied in the literature \cite{jafarkhani2001multiple,hochwald2000unitary,hassibi2002cayley}, it is also understood that the performance of DSTM deteriorates substantially in fast fading channels \cite{schober2002noncoherent,liu2004structures,nguyen2012performance}. Therefore, direct application of DSTM for HST communications is not a good option.

Motivated by this, we propose an improved DSTM scheme, which exploits the \emph{joint spatial and temporal channel correlation inherent in moving antenna array platforms}. At this point, it is worth emphasizing that the correlation effect considered in the current work differs significantly from the conventional concepts of \emph{spatial} or \emph{temporal} correlation. In the MIMO literature, see for instance \cite{kermoal2002stochastic,weichselberger2006stochastic}, the term \emph{spatial correlation} in general refers to \emph{antenna correlation} mainly due to the insufficient separation between the antenna elements. However, we assume that the antennas at both the transmitter and receiver are sufficiently separated, as such, there exists no \emph{antenna correlation} between antenna elements. Also, conventionally, the temporal correlation \cite{li1998robust,choi2001channel,ningsun2014spectral} is used to model the channel variation of the same channel at different time instants. Based on this terminology, coherence time and quasi-stationary channel were subsequently formulated. However, under high mobility conditions, the stationary interval of the channel is extremely small. Therefore, the joint spatial and temporal correlation model proposed in this paper is different from those appeared in prior works, it is used to model the relationship between channel realizations of distinct antenna pairs at different time instants due to the mobility of the antenna array. We will elaborate this point later in Section II. Leveraging this new correlation information, the proposed DSTM scheme is shown to achieve robust communications under high mobility conditions.

The main contributions of this paper are summarized as follows:
\begin{itemize}
    \item We discover a novel spatial and temporal correlation pattern inherent to wireless communications systems with moving antenna array. In addition, the correlation structure is derived, and the impact of key physical parameters on channel correlation is characterized.
    \item Capitalizing on the new correlation information, the optimal decoding criterion for the DSTM scheme is derived. In addition, pairwise error probability (PEP) upper-bounds are obtained, which enables the study of the impact of node mobility and transmission block length on the system performance. Analytical results demonstrates that exploiting channel correlation statistics brings a significant performance gain in high mobility settings.
    \item An adaptive transmission method, where the transmitter adjusts the block length adaptively according to the node velocity, is proposed. It was shown that the proposed adaptive scheme yields improved error performance.
\end{itemize}

The remainder of this paper is organized as follows. In Section II, the joint spatial and temporal channel correlation model is presented. An improved DSTM scheme and its error performance is studied in Section III. In Section IV, an adaptive transmission method is proposed for further performance improvement. Finally, Section V concludes the paper.

Throughout the rest of the paper, matrices and vectors are denoted by bold capital and lowercase letters, respectively. Let $\mathbf{V}\in\mathbb{C}^{M\times N}$ mean that the complex matrix $\mathbf{V}$ is consist of $M$ rows and $N$ columns. $\textbf{I}_{N}$ denotes an $N\times N$ identity matrix; $\textrm{E}\left[\cdot\right]$ is the mathematical expectation operator; $\left ( \cdot  \right )^{H}$ stands for Hermitian transpose; $\left\|\cdot\right\|_{\textrm{F}}$ refers to the Frobenius norm; $\mathrm{Re}\left\{\cdot\right\}$ gives the real part; $\mathrm{Tr}\left\{ \cdot \right\}$ is the matrix trace operation; $\left\lfloor \cdot \right\rfloor$ and $\left\lceil \cdot \right\rceil$ denote, respectively, the floor and ceiling functions; $\mathrm{Round}\left\{ \cdot \right\}$ rounds the element to the nearest integer.

\section{Joint Spatial and Temporal Channel Correlation Model}

\begin{figure}[ht]\centering
\includegraphics[angle=0,scale=0.1]{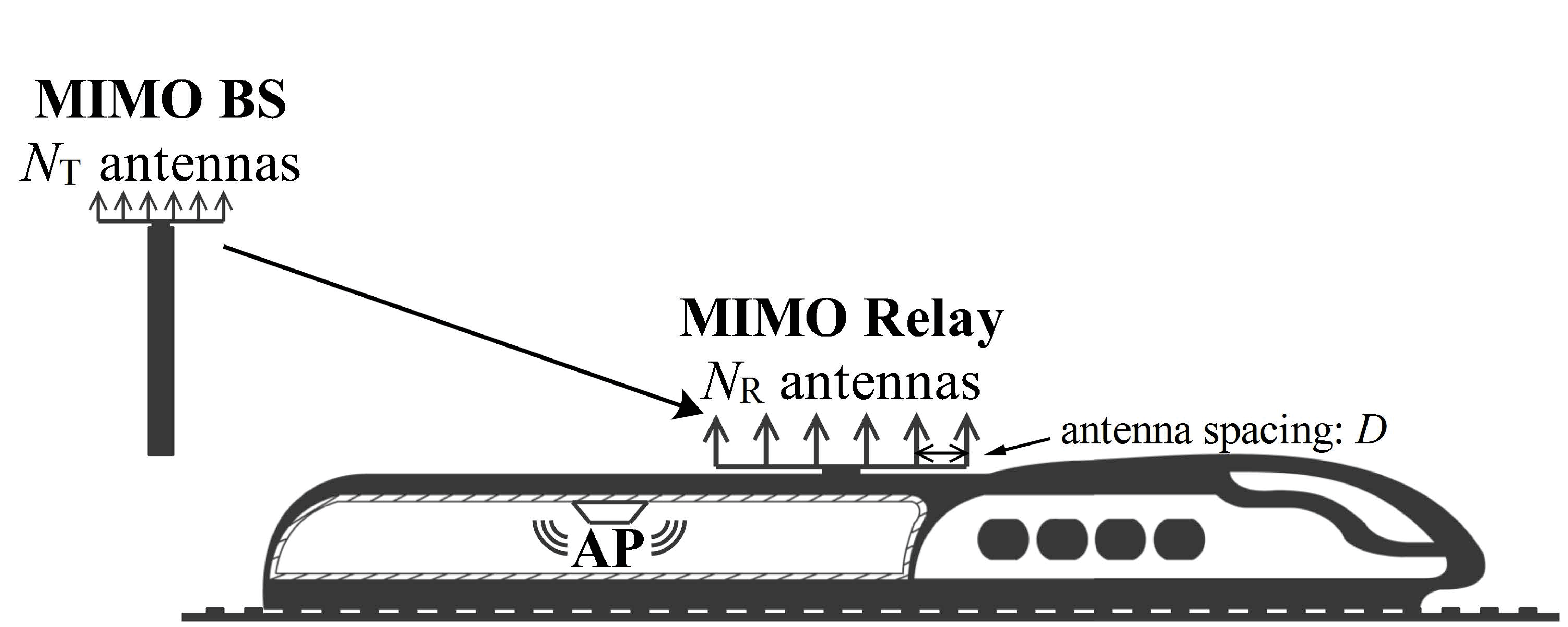}
\caption{Schematic model for the HST scenario.}
\label{Fig_Scenario}
\end{figure}

We consider the HST communication scenario\footnote{While this paper focuses on the HST scenario, the proposed correlation model also applies to other high mobility scenarios.}, where the base station (BS) equipped with $N_{\textrm{T}}$ antennas sends information to passengers with the help of a dedicated relay equipped with $N_{\textrm{R}}$ antennas with uniform spacing $D$ mounted on top of the train, as illustrated in Fig. \ref{Fig_Scenario}. Direct links between the BS and passengers do not exist due to heavy penetration loss of the radio signal through the well shield carriages \cite{3gpp2014mobile}. As such, access points (AP) are installed inside the train to provide network services for passengers. Since the communication links between the AP and passengers are almost static, establishing high speed connections is relatively simple. Hence, in this paper, we focus on the more challenging point-to-point communication link between the BS and relay.

\begin{figure}[ht]\centering
\includegraphics[angle=0,scale=0.15]{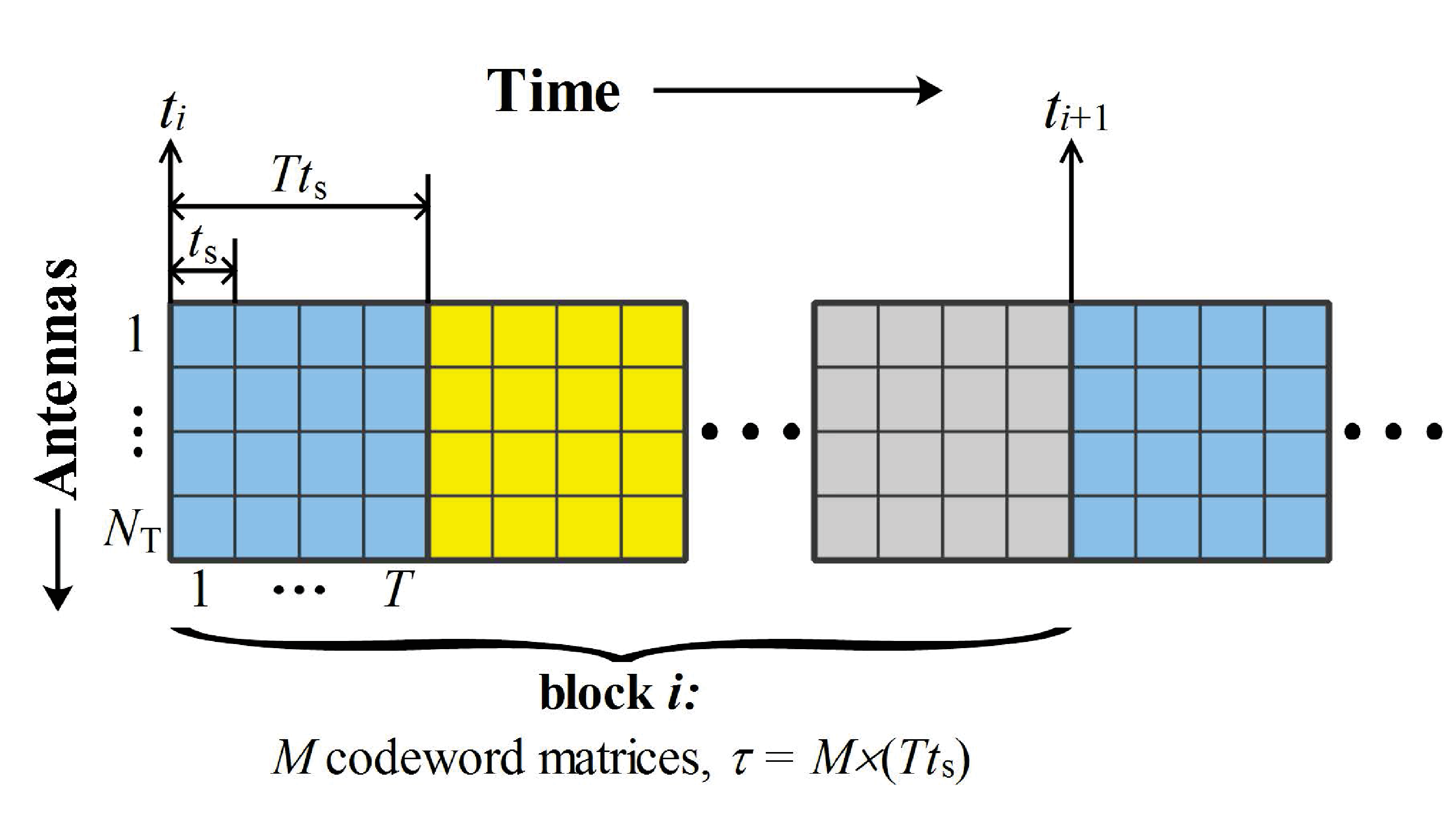}
\caption{Information transmission block.}
\label{Fig_Transmission}
\end{figure}

At time $t_{i}$, the BS transmits an information block (denoted as block $i$) containing $M$ codeword matrices of size $N_{\textrm{T}}\times T$, as illustrated in Fig. \ref{Fig_Transmission}. Now, denoting the $j$-th codeword matrix in block $i$ as $\textbf{X}_{i,j}$, then the $j$-th matrix observed at the receiver in the $i$-th block is given by
\begin{equation}
\textbf{Y}_{i,j}=\sqrt{P}\textbf{H}_{i,j}\textbf{X}_{i,j}+\textbf{N}_{i,j},
\label{Received-Symbol}
\end{equation}
where $\textbf{Y}_{i,j}\in \mathbb{C}^{N_{\textrm{R}} \times T}$ is the received codeword matrix, $\textbf{H}_{i,j}\in \mathbb{C}^{N_{\textrm{R}} \times N_{\textrm{T}}}$ is the channel matrix between the BS and Relay. $P$ is the transmit power of each antenna and $\textbf{N}_{i,j}\in \mathbb{C}^{N_{\textrm{R}} \times T}$ is the zero-mean complex Gaussian noise with each column having covariance matrix $\textbf{R}_{\textrm{n}}=\sigma^{2}_{\textrm{n}}\textbf{I}_{N_{\textrm{R}}}$.

In this paper, we assume that the entries of $\textbf{H}_{i,j}$ are independent and identically distributed (i.i.d.) complex Gaussian variables with zero-mean and unit-variance. Please note, such isotropic model is particularly relevant to the scenarios where the propagation environment is dominated by the diffuse counterpart, such as viaduct-1a and cuttings \cite{ai2012radio}. Moreover, we assume that the channel matrix $\textbf{H}_{i,j}$ remains static during one codeword duration $T t_{\textrm{s}}$ with $t_{\textrm{s}}$ being the symbol duration. To this end, it is required that $T t_{\textrm{s}} \leq t_{\textrm{c}}$, where $t_{\textrm{c}}\approx \frac{0.423}{f_{d}}$ is the coherence time and $f_{d}$ represents the maximum Doppler frequency.

\subsection{Channel Correlation Model}
We now give a detailed explanation on the channel correlation model adopted in the paper.

\begin{figure}[ht]\centering
\includegraphics[angle=0,scale=0.135]{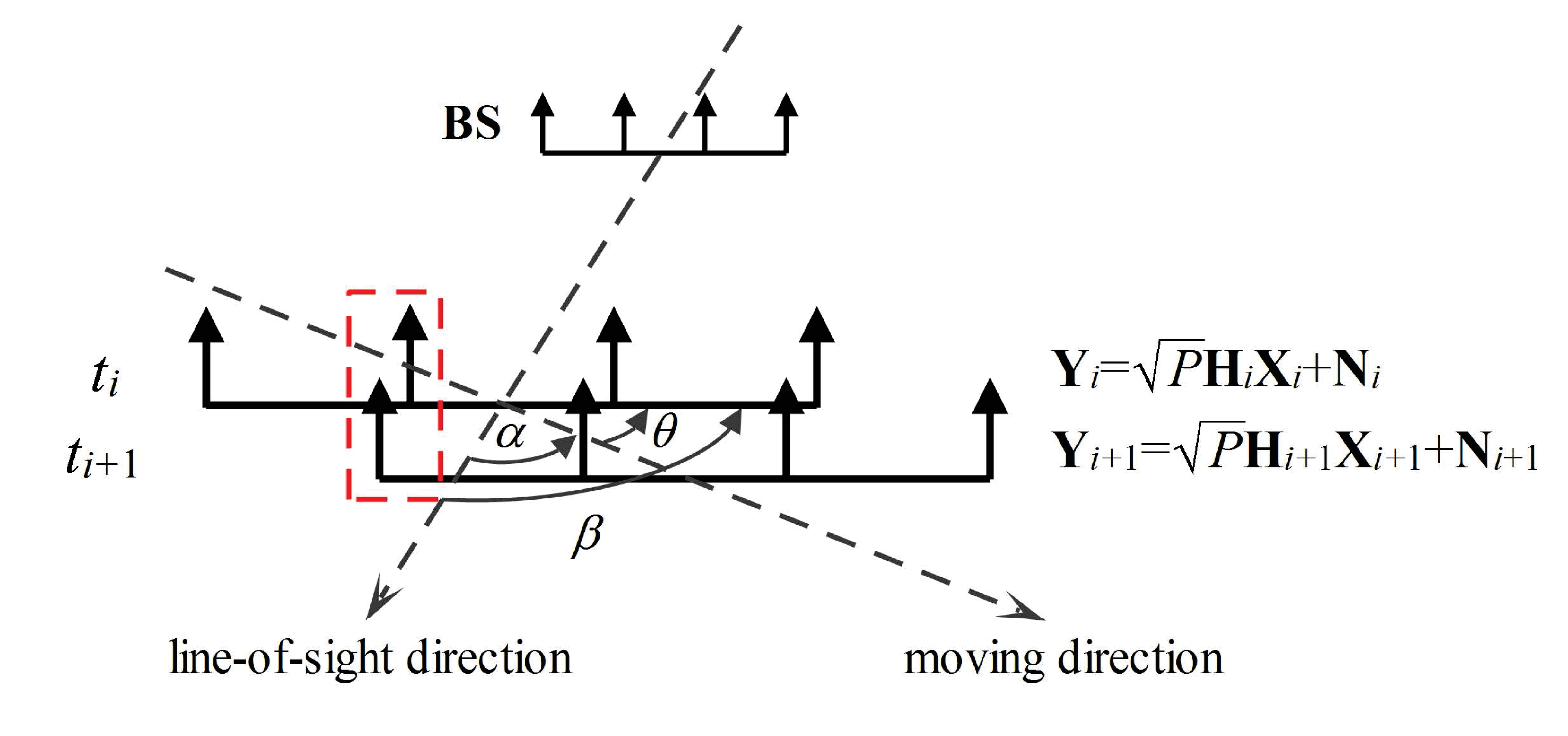}
\caption{Moving antenna array introduces joint spatial and temporal channel correlation.}
\label{Fig_Mobility_Model}
\end{figure}

Consider the moving antenna array depicted in Fig. \ref{Fig_Mobility_Model}. Without loss of generality, we assume that the HST has constant velocity $v \in [0,+\infty)$ and moving direction $\theta \in (-\frac{1}{2}\pi,\frac{1}{2}\pi)$. Intuitively, once $v$ and $\theta$ are specified, the relative position between the antennas at $t_{i+1}$ and $t_{i}$ is determined. Throughout this paper, we assume that $v$ and $\theta$ are perfectly known by the HST. This is reasonable with the aid of mature tracking and positioning techniques such as the Global Positioning System (GPS).

Now let us consider a special case in which the left most antenna at time $t_{i+1}$ arrives at almost the same physical location of the second left most antenna at time $t_i$. Assume that the positions of these two antennas are within half wavelength. Then, combined with the fact that $\tau$ is very short due to high mobility, it can be inferred that the scattering environment remains almost identical. In this sense, it is very likely that the channel between the BS and the first antenna at $t_{i+1}$ is highly correlated with the channel between the BS and the second antenna at $t_i$. With an abuse of terminology, we refer this type of correlation as \emph{spatial-temporal correlation}.\footnote{The observation of the ``spatial-temporal channel correlation'' phenomenon has been independently made in a recent work \cite[Fig. 2]{Nirwan}, where a novel antenna adapting method was proposed to exploit this desirable feature to address the rapid channel variations in HST communication systems.}

Noticing the fact that such spatial-temporal correlation only exists between different codeword blocks, i.e., between channel matrices ${\bf H}_{i,j}$ and ${\bf H}_{i+1,j}$, we drop the subscript $j$ for notational convenience. Now, expressing the channel matrix ${\bf H}_{i}$ into column vector form
\begin{equation}
\textbf{H}_{i}=[\textbf{h}_{1,i}~\textbf{h}_{2,i}~\dots~\textbf{h}_{N_{\textrm{T}},i}],
\label{Channel-Decomposition}
\end{equation}
it is easy to see that the correlation structure between $\textbf{h}_{m,i}$ and $\textbf{h}_{m,i+1}$ is the same as that between $\textbf{h}_{n,i}$ and $\textbf{h}_{n,i+1}$, i.e., $\textrm{E}[\textbf{h}_{m,i+1}\textbf{h}_{m,i}^{H}]=\textrm{E}[\textbf{h}_{n,i+1}\textbf{h}_{n,i}^{H}],~1\leq m,n \leq N_{\textrm{T}}$. Denoting the correlation matrix $\textbf{C}$ as
\begin{equation}
\textbf{C}=\textrm{E}[\textbf{h}_{m,i+1}\textbf{h}_{m,i}^{H}],~1\leq m \leq N_{\textrm{T}},
\label{Correlation-Model}
\end{equation}
then invoking the vector autoregressive (VAR) model \cite{reinsel1992vector,hamilton1994time}, the relation between $\textbf{h}_{m,i+1}$ and $\textbf{h}_{m,i}$ is formulated by
\begin{equation}
\textbf{h}_{m,i+1}=\textbf{C}\textbf{h}_{m,i}+\textbf{u}_{m,i+1},~1\leq m \leq N_{\textrm{T}},
\label{AR1-Model}
\end{equation}
where $\textbf{u}_{m,i+1}\in \mathbb{C}^{N_{\textrm{R}}\times1}$ is a random complex Gaussian noise vector with covariance matrix $\textbf{R}_{\textrm{u}}$. To ensure that $\left\{ \textbf{h}_{m,i} \right\}_{i}$ is a stationary VAR(1) time series, we need $\textbf{R}_{\textrm{u}}=\textbf{I}_{N_{\textrm{R}}}-\textbf{C}\textbf{C}^{H}$.

Combining (\ref{Channel-Decomposition}), (\ref{Correlation-Model}) and (\ref{AR1-Model}), the correlation between channel samples can be rewritten in a compact matrix form as
\begin{equation}
\textbf{H}_{i+1}=\textbf{C}\textbf{H}_{i}+\textbf{U}_{i+1},
\label{Overall-AR1-Model}
\end{equation}
where $\textbf{U}_{i+1}\triangleq[\textbf{u}_{1,i+1}~\textbf{u}_{2,i+1}~\dots~\textbf{u}_{N_{\textrm{T}},i+1}]$.

\newcounter{TempEqCnt1}
\setcounter{TempEqCnt1}{\value{equation}}
\setcounter{equation}{14}
\begin{figure*}[ht]

\begin{equation}
\rho \left ( \tau ,D \right ) = \frac{I_{0}\left( \sqrt{ \kappa^2-a^{2}-b^{2}+2ab\cos\left(\beta - \alpha \right )-j2\kappa \left[ a\cos\left( \mu - \alpha \right)-b\cos\left( \mu-\beta \right ) \right ] } \right)}{I_{0}\left( \kappa \right)}e^{-c_{0}v\left | \tau \right |},
\label{modi_correlation_model}
\end{equation}
\hrulefill
\end{figure*}
\setcounter{equation}{\value{TempEqCnt1}}

\subsection{Structure of the Correlation Matrix ${\bf C}$}
We now study the structure of the correlation matrix $\textbf{C}$. Before going into the details, we first state the following important fact which will be invoked in the ensuing derivations.
\begin{fact}
If the distance between the two antennas are sufficiently large, then there is no spatial-temporal correlation between them.
\end{fact}
This Fact is rather intuitive, and has been adopted in many channel models. For instance, according to the Clark's model, antennas separated by a distance over half wavelength can be regarded as spatially uncorrelated. In the current work, we generalize this Fact to account for the correlation effect between antenna pairs at different time. For instance, the distance between the first left most antenna at time $t_{i}$ and the second left most antenna at time $t_{i+1}$ is larger than half wavelength, hence, there is no spatial-temporal correlation.

Keeping Fact 1 in mind, we are ready to present the following key results:
\begin{lemma}\label{lemma1}
The $N_{\emph{R}} \times N_{\emph{R}}$ correlation matrix ${\bf C}$ can be modeled by a Toeplitz matrix as follows
\begin{equation}
\emph{\textbf{C}}=
\begin{bmatrix}
 \cdots &\rho_{l}  &\rho_{l+1}   &        &\emph{\large{\textbf{0}}}        \\
        &        &\rho_{l}  &\ddots  &        \\
        &        &        &\ddots  &\rho_{l+1}   \\
        &\emph{\large{\textbf{0}}}      &        &        &\rho_{l}  \\
        &        &        &        & \vdots
\end{bmatrix}
,
\label{Toeplitz-correlation-Model}
\end{equation}
where $l$ is some context-specific integer, while $\rho_{l}$ and $\rho_{l+1}$ denote the correlation coefficients, and they are determined by system parameters including block duration $\tau$, velocity $v$, antenna spacing $D$, and moving direction $\theta$, etc.
\end{lemma}
\begin{IEEEproof}[Proof]
See Appendix \ref{proof_lemma1}.
\end{IEEEproof}

\emph{Remarks}: Lemma \ref{lemma1} indicates that at time $t_{i+1}$, the channel gain $h_{m,n,i+1}$ is at most correlated with two entries of $\textbf{h}_{m,i}$, with correlation coefficients $\rho_{l}$ and $\rho_{l+1}$, respectively. This observation is rather intuitive, and is a direct consequence of Fact 1.

\begin{figure}[h!t]
\centering
\subfigure[Case I: $\theta=0$, $v\tau=0.5D$]{
\label{caseI} 
\includegraphics[angle=0,scale=0.22]{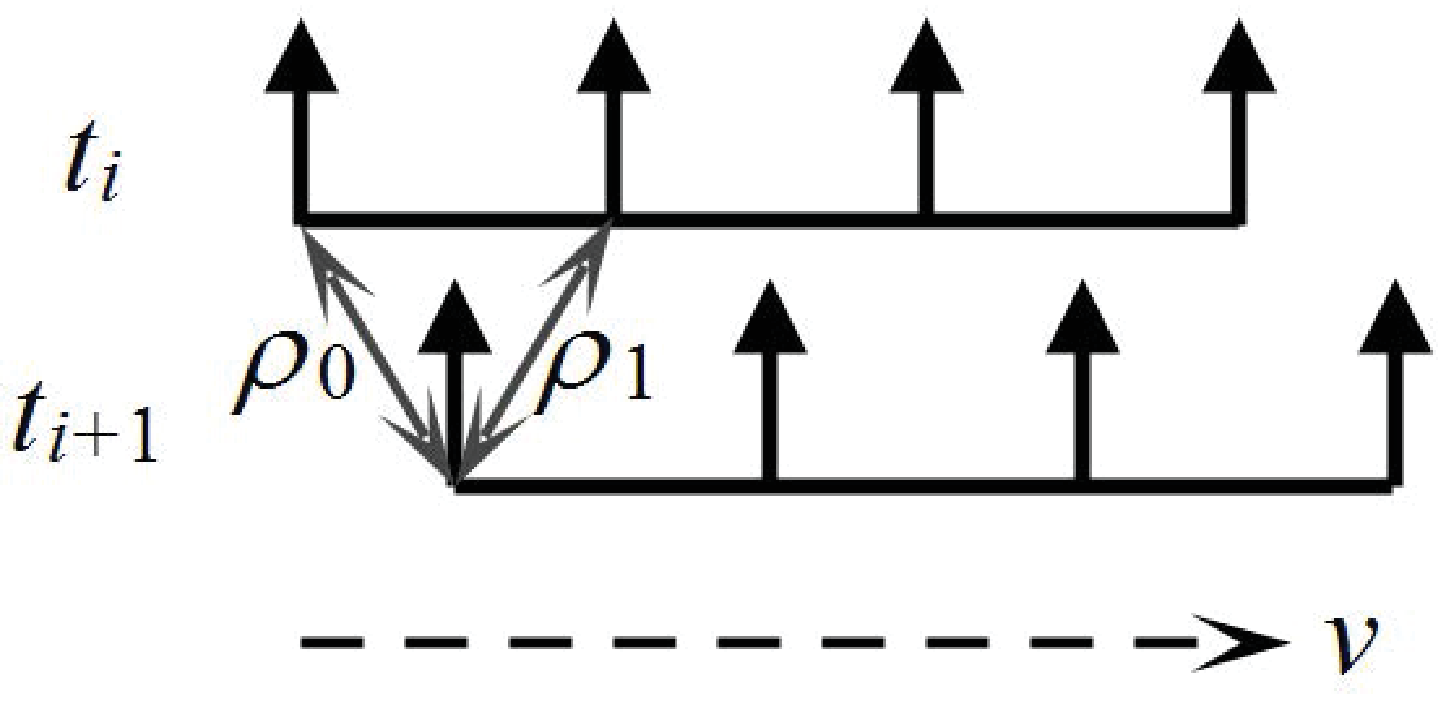}}
\hspace{0.15in}
\subfigure[Case II: $\theta=0$, $v\tau=D$]{
\label{caseII} 
\includegraphics[angle=0,scale=0.22]{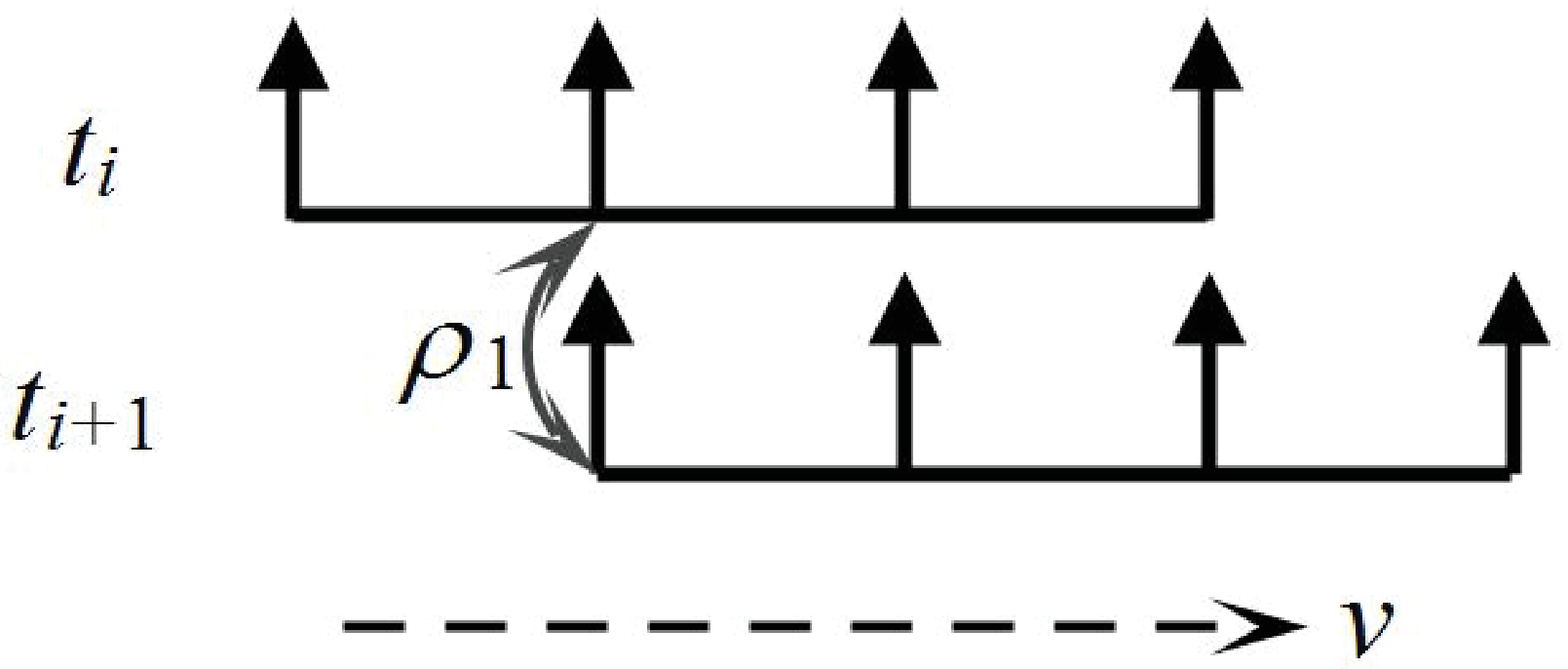}}
\hspace{0.15in}
\subfigure[Case III: $\theta=0.5\pi$, $v\tau>0.5\lambda$]{
\label{caseIIIa} 
\includegraphics[angle=0,scale=0.22]{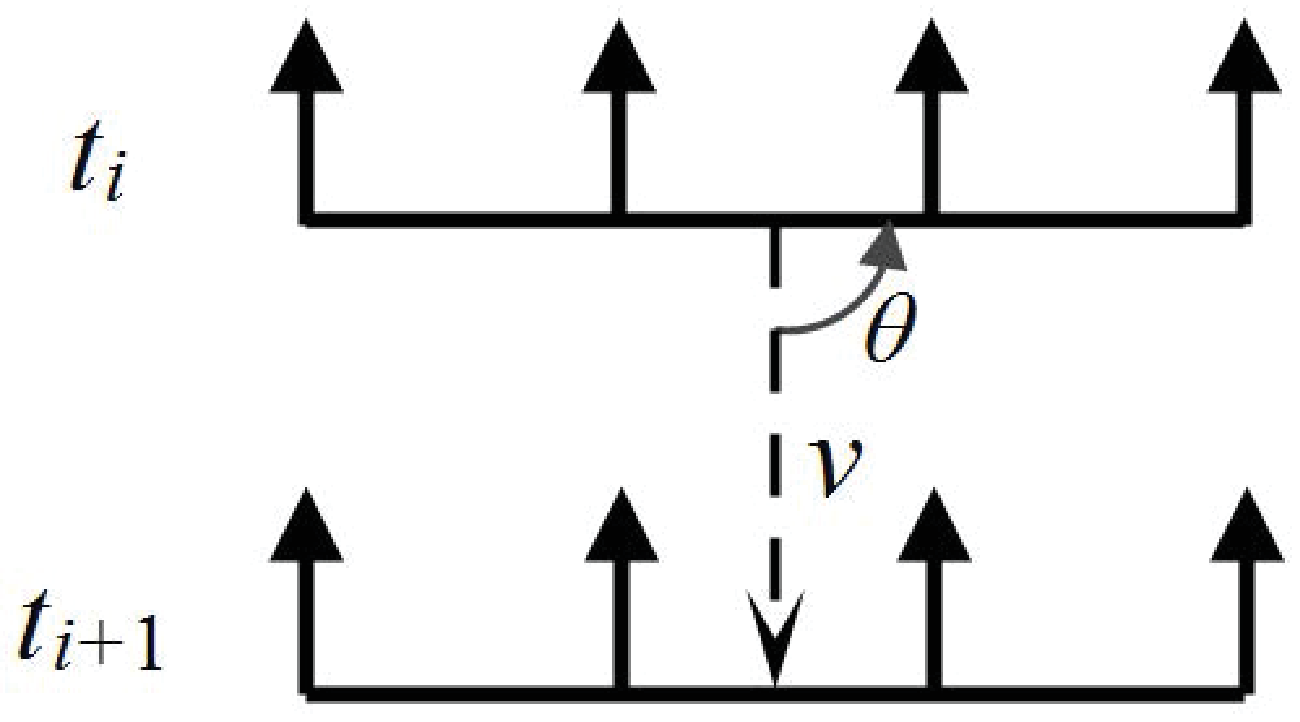}}
\hspace{0.15in}
\subfigure[Case III: $\theta=0$, $v\tau=4D$]{
\label{caseIIIb} 
\includegraphics[angle=0,scale=0.22]{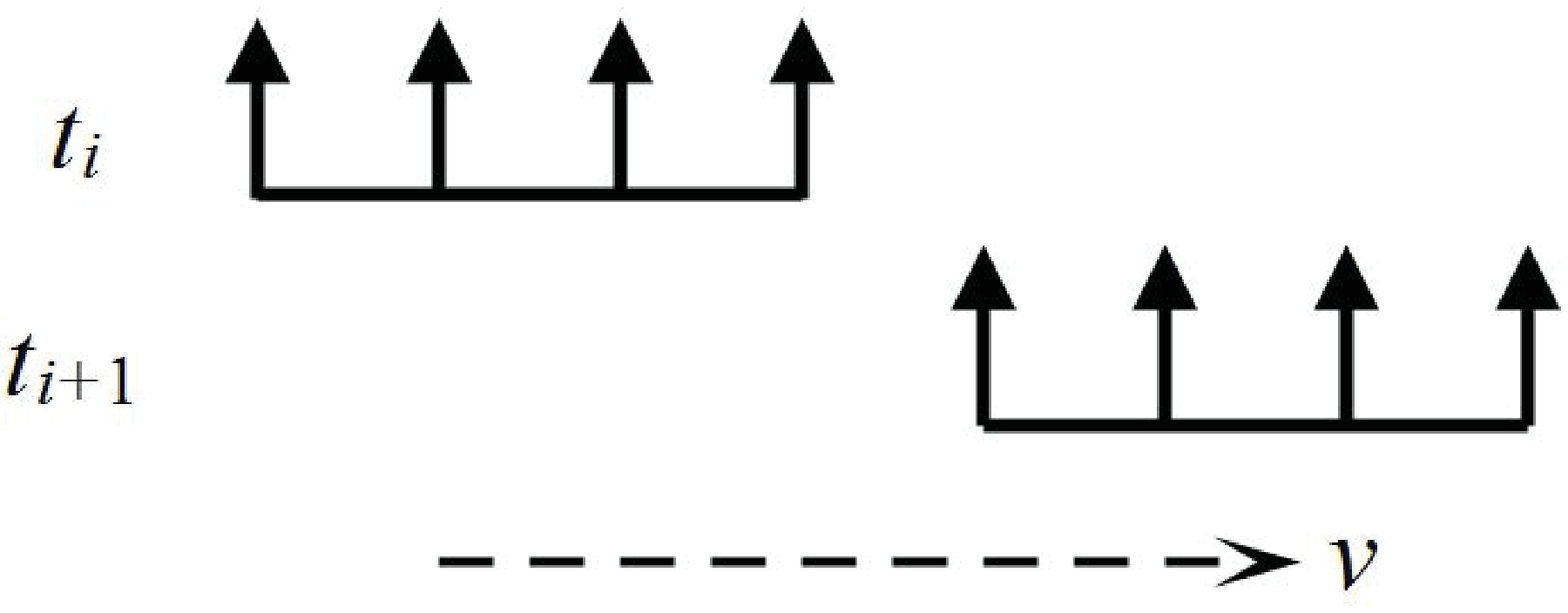}}
\caption{Specific examples for the given three cases.}
\label{cases} 
\end{figure}

Without loss of generality, assume that the distance threshold in Fact 1 is $0.5\lambda$. Then, based on some simple geometry analysis of the spatial-temporal model depicted in Fig. \ref{Fig_Mobility_Model}, the matrix $\textbf{C}$ may exhibit three different structure and properties:

\emph{Case I}: There exists an $l$ that satisfies the following conditions
\begin{equation}
\sqrt{\left( v\tau \cos(\theta)-lD \right)^{2}+\left(v\tau \sin(\theta)\right)^{2}}\leq 0.5\lambda, \mbox{ and}
\label{Correlation_case_a_condition1}
\end{equation}
\begin{equation}
\sqrt{\left( lD+D-v\tau \cos(\theta) \right)^{2}+\left(v\tau \sin(\theta)\right)^{2}}\leq 0.5\lambda.
\label{Correlation_case_a_condition2}
\end{equation}
In such scenarios, the position of the first antenna at $t_{i+1}$, is within half wavelength of the $(l+1)$-th and the $(l+2)$-th antenna at $t_{i}$, implying strong correlation between the corresponding channels as illustrated in Fig. \ref{cases}(a). Based on Fact 1, it can be inferred that
\begin{equation}
\rho_{l}\neq 0,~\rho_{l+1}\neq 0, \mbox{ and }l=\lfloor \frac{v\tau \cos( \theta)}{D} \rfloor.
\label{Correlation_case_a}
\end{equation}

\emph{Case II}: There exists an $l$ that satisfies the following conditions
\begin{equation}
\sqrt{\left( v\tau \cos(\theta)-lD \right)^{2}+\left(v\tau \sin(\theta)\right)^{2}}\leq 0.5\lambda, \mbox{ and}
\label{Correlation_case_b_condition1}
\end{equation}
\begin{equation}
\sqrt{\left( v\tau \cos(\theta)-lD+D \right)^{2}+\left(v\tau \sin(\theta)\right)^{2}}> 0.5\lambda, \mbox{ and}
\label{Correlation_case_b_condition2}
\end{equation}
\begin{equation}
\sqrt{\left( lD+D-v\tau \cos(\theta) \right)^{2}+\left(v\tau \sin(\theta)\right)^{2}}> 0.5\lambda.
\label{Correlation_case_b_condition3}
\end{equation}
In such scenarios, the channel of the first antenna at $t_{i+1}$ is only correlated to the $(l+1)$-th antenna at $t_{i}$ as illustrated in Fig. \ref{cases}(b). Hence, we have
\begin{equation}
\rho_{l}\neq 0, \mbox{ and }l=\mathrm{Round}\left\{ \frac{v\tau \cos( \theta)}{D} \right\}.
\label{Correlation_case_b}
\end{equation}

\emph{Case III}: There exists no such $l$ that satisfies
\begin{equation}
\sqrt{\left( v\tau \cos(\theta)-lD \right)^{2}+\left(v\tau \sin(\theta)\right)^{2}}\leq 0.5\lambda,
\label{Correlation_case_c_condition}
\end{equation}
which indicates that the channel at $t_{i+1}$ is almost uncorrelated to that at $t_{i}$, as illustrated in Fig. \ref{cases}(c) and Fig. \ref{cases}(d). Equivalently, in such scenarios $\textbf{C}$ is approximately all-zero.

\subsection{A Specific Model for Correlation Coefficients}

In general, the correlation coefficients $\rho_{l}$ and $\rho_{l+1}$ can be estimated from measurement data. However, we may exploit a specific spatial-temporal correlation structure to establish an analytical model. Building upon the channel correlation framework proposed in \cite{abdi2002space}, the moving scattering objects are modeled as poisson point processes (PPP), which results in an exponential decay factor attached to the cross-correlation function. Hence, the final correlation coefficient expression can be obtained as (\ref{modi_correlation_model}), where $I_{0}( z )$ is $0$-th order modified Bessel function of the first kind. $\alpha$, $\beta$ are defined in Fig. \ref{Fig_Mobility_Model}, and $\theta = \beta - \alpha$. Moreover, $a\triangleq 2\pi f_{D}\tau$ and $b\triangleq 2\pi D/\lambda$ where $f_{D}=\frac{v}{\lambda}$ is the maximum Doppler shift and $\lambda$ is the carrier wavelength. The parameter $\kappa$ indicates the width of angle of arrival (AOA) and $\mu\in\left[-\pi, \pi\right)$ accounts for the mean direction of AOA. $c_{0}$ is a real constant characterizing the spatial property of the scattering objects. For the special case of isotropic scattering, we have $\kappa=0$ and $\tau >0$, as such, $\rho_{l}$ and $\rho_{l+1}$ can be expressed as
\setcounter{equation}{15}
\begin{equation}
\rho_{k}=
\left\{\begin{matrix}
J_{0}\left( \sqrt{a^{2}+b_{k}^{2}-2ab_{k}\cos\left( \theta \right)} \right)e^{-c_{0}v \tau },~k=l,l+1
\\
0,~~~~~~~\mbox{otherwise}~~~~~~~~~~~~
\end{matrix}\right.
\label{alpha_and_beta}
\end{equation}
where $J_{0}\left( z \right)$ is $0$-th order Bessel function of the first kind and $b_{k}\triangleq 2\pi kD/\lambda$. Note that a similar channel correlation model has been adopted in \cite{jakes1994microwave} and \cite{truong2013effects}.

As mentioned above, $v$ and $\theta$ are known as priori information at the HST. Then $\rho_{l}$ and $\rho_{l+1}$ can be directly derived from (\ref{alpha_and_beta}). Invoking Lemma $1$, this indicates that $\textbf{C}$ can be obtained by the relay.

\section{A Novel Differential Modulation Scheme}
In this section, we first propose a novel differential modulation scheme to exploit the spatial-temporal correlation structure discovered in the previous section. Then, the error performance of the proposed scheme is studied by deriving an uniform PEP upper bound. Finally, we look at the special case when the correlation matrix $\textbf{C}$ has only one superdiagonal or main diagonal.

\newcounter{TempEqCnt2}
\setcounter{TempEqCnt2}{\value{equation}}
\setcounter{equation}{24}
\begin{figure*}[ht]
\begin{equation}
\textrm{Pr}\left \{ \textbf{G}\rightarrow {\textbf{G}}' \right \}\leq
\frac{1}{\left | \textbf{I}_{N_\textrm{T}}+\frac{\gamma^{2}}{1+2\gamma }\left [ \textbf{I}_{N_\textrm{T}}-\frac{1}{ 4N_{\textrm{T}}}\textbf{D}\left ( \textbf{I}_{N_{\textrm{T}}}+{\textbf{G}}'\textbf{G}^{H} \right )\left ( \textbf{I}_{N_{\textrm{T}}}+\textbf{G}{\textbf{G}}'^{H} \right )\textbf{D}^{H} \right ] \right |^{N_{\textrm{R}}-l}},
\label{PEP}
\end{equation}
\hrulefill
\end{figure*}
\setcounter{equation}{\value{TempEqCnt2}}

\subsection{Proposed Differential Modulation Scheme}
Space-time differential modulation schemes based on orthogonal designs have been extensively studied in the MIMO literature, see for instance \cite{hochwald2000unitary,hughes2000differential,hassibi2002cayley}, where the codeword matrix $\textbf{X}_{i}$ satisfies
the following orthogonality properties:
\begin{equation}
\textbf{X}_{i}\textbf{X}_{i}^H=T\textbf{I}_{N_{\textrm{T}}}, \mbox{ and } \textbf{X}_{i}^H\textbf{X}_{i}=N_{\textrm{T}}\textbf{I}_{T}.
\label{Orthogonal-Design}
\end{equation}

However, the conventional differential modulation schemes are designed for static or quasi-static channels, therefore can not handle the fast fading channel scenarios under high mobility conditions. To tackle this critical issue, we propose an improved differential space-time modulation scheme, exploiting the spatial-temporal correlation structure of the moving antenna array.

For differential transmission, the unitary codeword matrices $\textbf{X}_{0},\textbf{X}_{1},\dots,\textbf{X}_{i},\dots$ are organized in a way similar to DPSK:
\begin{equation}
\begin{aligned}
&\textbf{X}_{0}=\textbf{D},\\
&\textbf{X}_{i+1}=\textbf{X}_{i}\textbf{G}_{k},~\textbf{G}_{k}\in\mathcal{G},~k=1,\dots,K
\end{aligned}\label{Differential-encoding}
\end{equation}
where $\textbf{D}$ is the initial matrix known \emph{a priori} at the receiver, and satisfies $\textbf{D}\textbf{D}^{H}=\textbf{D}^{H}\textbf{D}=N_{\textrm{T}}\textbf{I}_{N_{\textrm{T}}}$ according to (\ref{Orthogonal-Design}), and $\textbf{G}_{k}$ is the information-bearing matrix which belongs to an $N_{\textrm{T}}\times N_{\textrm{T}}$ unitary matrices set $\mathcal{G}$ with cardinality $K$ and is carefully designed to ensure that $\textbf{X}_{i+1}\in \textbf{D}\mathcal{G}$ whenever $\textbf{X}_{i}\in \textbf{D}\mathcal{G}$. Accordingly, $\textbf{D}\mathcal{G}=\left\{ \textbf{D}\textbf{G}:\textbf{G}\in \mathcal{G} \right\}$ is referred to as the \emph{differential code set}.

We now look at the differential detection phase. Substituting (\ref{Overall-AR1-Model}) and (\ref{Differential-encoding}) into (\ref{Received-Symbol}), the received symbol matrices at time $t_i$ and $t_{i+1}$ can be expressed as
\begin{equation}
\begin{split}
\textbf{Y}_{i}~=~&\sqrt{P}\textbf{H}_{i}\textbf{X}_{i}+\textbf{N}_{i}, \mbox{ and }\\
\textbf{Y}_{i+1}~=~&\sqrt{P}\textbf{H}_{i+1}\textbf{X}_{i+1}+\textbf{N}_{i+1}\\
                ~=~&\sqrt{P}\textbf{C}\textbf{H}_{i}\textbf{X}_{i+1}+\sqrt{P}\textbf{U}_{i+1}\textbf{X}_{i+1}+\textbf{N}_{i+1}\\
                ~=~&\sqrt{P}\textbf{C}\textbf{H}_{i}\textbf{X}_{i}\textbf{G}_{k}+\sqrt{P}\textbf{U}_{i+1}\textbf{X}_{i}\textbf{G}_{k}+\textbf{N}_{i+1}.
\label{Two-Received}
\end{split}
\end{equation}

Invoking the analysis in \cite{hughes2000differential}, we have the following theorem.

\begin{theorem}\label{theorem1}
The optimal differential decoding criterion is given by
\begin{equation}
\widehat{\textbf{\emph{G}}}=\mathrm{arg}\underset{\textbf{\emph{G}}\in \mathcal{G}}{\mathrm{max}}\mathrm{Re}\left \{ \mathrm{Tr} \left \{\textbf{\emph{G}}\textbf{\emph{Y}}_{i+1}^{H}\textbf{\emph{C}}\textbf{\emph{Y}}_{i} \right \}\right \}.
\label{Improved-Decoding}
\end{equation}
\end{theorem}
\begin{IEEEproof}[Proof]
See Appendix \ref{proof_theorem1}.
\end{IEEEproof}

For better understanding, we give (\ref{Improved-Decoding}) an \emph{estimator-correlator} interpretation as for the conventional maximum likelihood (ML) differential decoder defined by \cite{hughes2000differential}
\begin{equation}
\widehat{\textbf{G}}~=~\mathrm{arg}\underset{\textbf{G}\in \mathcal{G}}{\mathrm{max}}\mathrm{Re}\left\{ \mathrm{Tr} \left \{\textbf{G}\textbf{Y}_{i+1}^{H}\textbf{Y}_{i} \right \}\right\}.
\label{Standard-Decoding}
\end{equation}
It can be observed that (\ref{Standard-Decoding}) is just a special case of (\ref{Improved-Decoding}) when ${\bf C} = {\bf I}$ and $\textbf{U}_{i} = \bf 0$. For this case, since $\textbf{H}_{i+1}= \textbf{H}_{i}$, $\textbf{Y}_{i+1}$ and $\textbf{Y}_{i}$ can be used to estimate $\sqrt{P}\textbf{H}_i\textbf{X}_{i+1}$ (=$\sqrt{P}\textbf{H}_{i+1}\textbf{X}_{i+1}$) and $\sqrt{P}\textbf{H}_{i}\textbf{X}_{i}$, respectively. Hence, the decoding of $\textbf{G}$ can be completed through maximizing the correlation between $\textbf{Y}_{i+1}$ and $\textbf{Y}_{i}\textbf{G}$.

However, in our studied scenario, since now $\textbf{H}_{i+1} \neq \textbf{H}_{i}$, albeit $\sqrt{P}\textbf{H}_{i}\textbf{X}_{i}$ can still be directly estimated by $\textbf{Y}_{i}$, we need to find an alternate estimate for $\sqrt{P}\textbf{H}_{i}\textbf{X}_{i+1}$ instead of $\textbf{Y}_{i+1}$ in order to decode $\textbf{G}$. According to (\ref{Two-Received}), a column of $\textbf{Y}_{i+1}$ can be expressed as
\begin{equation}
\textbf{y}_{i+1}=\sqrt{P}\textbf{C}\textbf{H}_{i}\textbf{x}_{i+1}+\sqrt{P}\textbf{U}_{i}\textbf{x}_{i+1}+\textbf{n}_{i+1},
\label{Receive-subblock}
\end{equation}
where $\textbf{x}_{i+1}$ and $\textbf{n}_{i+1}$ are the corresponding columns of $\textbf{X}_{i+1}$ and $\textbf{N}_{i+1}$, respectively. Recall that $\textrm{E}[\textbf{H}_{i}\textbf{H}_{i}^{H}]=N_{\textrm{T}}\textbf{I}_{N_{\textrm{R}}}$, $\textrm{E}[\textbf{U}_{i}\textbf{U}_{i}^{H}]=N_{\textrm{T}}(\textbf{I}_{N_{\textrm{R}}}-\textbf{C}\textbf{C}^{H} )$ and $\textrm{E}[\textbf{x}_{i+1}\textbf{x}_{i+1}^{H}]=\textbf{I}_{N_{T}}$, $\textbf{H}_{i}\textbf{x}_{i+1}$ can then be estimated by invoking the minimum mean square error (MMSE) estimator
\begin{align}
&\widehat{\textbf{H}_{i}\textbf{x}_{i+1}}\notag\\
&=\frac{N_{\textrm{T}}}{\sqrt{P}}\textbf{C}^{H}\left( N_{\textrm{T}}\left( \textbf{C}\textbf{C}^{H} + \textbf{I}_{N_{\textrm{R}}}- \textbf{C}\textbf{C}^{H} \right) +\frac{\sigma^{2}_{\textrm{n}}}{P}\textbf{I}_{N_{\textrm{R}}} \right)^{-1}\textbf{y}_{i+1}\notag\\
&=\frac{\sqrt{P} N_{\textrm{T}}}{ P N_{\textrm{T}}+\sigma^{2}_{\textrm{n}} } \textbf{C}^{H}\textbf{y}_{i+1}.
\label{Transmit-subblock-Estimation}
\end{align}
Therefore, the optimal estimate of $\textbf{H}_{i}\textbf{X}_{i+1}$ can be written as
\begin{equation}
\widehat{\textbf{H}_{i}\textbf{X}_{i+1}}=\frac{\sqrt{P} N_{\textrm{T}}}{ P N_{\textrm{T}}+\sigma^{2}_{\textrm{n}} }\textbf{C}^{H}\textbf{Y}_{i+1}.
\label{Block-Estimation}
\end{equation}
Thus replacing $\textbf{Y}_{i+1}$ in (\ref{Standard-Decoding}) by $\textbf{C}^{H}\textbf{Y}_{i+1}$ results in the proposed decoder (\ref{Improved-Decoding}).

\subsection{Uniform PEP Upper-Bound}

We now study the error performance of the proposed differential decoding scheme, and obtain an upper-bound for the PEP. By extending the Chernoff bound of the PEP of the ML receiver for the conventional unitary space-time coding system in a quasi-static channel (cf.\cite{hughes2000differential}, eq.(9)) to our studied channel with the specific spatial-temporal correlation, we have the following result:

\begin{theorem}\label{theorem2}
An upper bound of the PEP of the ML receiver for the proposed differential decoding scheme is given by (\ref{PEP}), where $\gamma$ is the equivalent signal-to-interference-plus-noise ratio (SINR) defined as
\setcounter{equation}{25}
\begin{equation}
\gamma=\frac{ \left ( 1-\frac{l}{N_{\textrm{R}}} \right )\left | \rho_{l} \right |^{2}+\left ( 1-\frac{l+1}{N_{\textrm{R}}} \right )\left | \rho_{l+1} \right |^{2} }
{ 1- \left ( 1-\frac{l}{N_{\textrm{R}}} \right )\left | \rho_{l} \right |^{2}-\left ( 1-\frac{l+1}{N_{\textrm{R}}} \right )\left | \rho_{l+1} \right |^{2} +\frac{\sigma^{2}_{\textrm{n}}}{N_{\textrm{T}}P} }.
\label{SINR}
\end{equation}
\end{theorem}
\begin{IEEEproof}[Proof]
See Appendix \ref{proof_theorem2}.
\end{IEEEproof}

From Theorem \ref{theorem2}, we can get the following important observations:

\emph{a)} Increasing the number of antennas at both the BS and the mobile receiver generally improves the error performance. This is a direct consequence of the facts that increasing $N_{\textrm{T}}$ and $N_{\textrm{R}}$ improves the SINR (as shown in (\ref{SINR})) and the PEP is a monotonically decreasing function with respect to $\gamma$.


\emph{b)} According to (\ref{Correlation_case_b}), if the train tends to slow down, i.e., $v\rightarrow0$, then $l\rightarrow0$. Hence, from (\ref{alpha_and_beta}) and (\ref{SINR}), there is $\rho_{0}\rightarrow1$, $\rho_{1}\rightarrow0$ and $\gamma\rightarrow\frac{N_{\textrm{T}}P}{\sigma^{2}_{\textrm{n}}}$. In such a case, (\ref{PEP}) reduces to the original PEP bound for differential unitary space-time codes (see (\ref{Original-PEP}) in Appendix \ref{proof_theorem2}), which indicates that the proposed scheme attains the same error performance as the conventional scheme.

\emph{c)} In the large $N_{\textrm{R}}$ and small $l$ regime, i.e., $\frac{l}{N_{\textrm{R}}}\rightarrow 0$, (\ref{SINR}) simplifies to $\gamma=\frac{ \left | \rho_{l} \right |^{2}+\left | \rho_{l+1} \right |^{2} }
{ 1- \left( \left| \rho_{l} \right|^{2}+\left | \rho_{l+1} \right|^{2} \right) +\frac{\sigma^{2}_{\textrm{n}}}{N_{\textrm{T}}} }$, which suggests that the PEP is a decreasing function with respect to $\left | \rho_{l} \right |^{2}+\left | \rho_{l+1} \right |^{2}$. Since in effect $\left | \rho_{l} \right |^{2}+\left | \rho_{l+1} \right |^{2}$ well reflects the degree of correlation between successive channel realizations, it indicates that the proposed scheme achieves superior error performance when the channel correlation is high.

To validate the aforementioned key observations, we now perform Monte Carlo simulations to illustrate the SER performance of the proposed scheme. For differential unitary space-time modulation, we take the optimal $4\times 4$ cyclic unitary group code $\mathcal{G}=\left \{ \textbf{I}_{4},\textbf{G},\textbf{G}^{2},\textbf{G}^{3} \right \}$ with a Hadamard initial matrix $\textbf{D}$ (Interested readers are referred to \cite{hughes2003optimal} for details regarding the optimal differential unitary space-time modulations). Hence, here $N_{\textrm{T}}=T=4$. Simulation parameters are summarized in Table I.

\begin{table}[htbp]
\begin{minipage}{0.45\textwidth}
\begin{center}
\caption[]{Simulation parameters}
\begin{tabular}{|c|c|}
\hline
Number of BS antennas $N_{\textrm{T}}$&$4$\\ \hline
Number of HST antennas $N_{\textrm{R}}$&$4$\\ \hline
Length of a codeword matrix $T$&$4$\\ \hline
SNR per receive antenna $=\frac{N_{\textrm{T}}P}{\sigma^{2}_{\textrm{n}}}$&$5$dB\\ \hline
Symbol duration $t_{\textrm{s}}$&$50\upmu$s\\ \hline
Spatial parameter $c_{0}$&$0.1$\\ \hline
Carrier frequency $f_{c}$&$3$GHz\\ \hline
Antenna spacing $D$&$0.5\lambda=5\times 10^{-2}$m\\ \hline
Number of codeword matrices in a block $M$&$1$\\ \hline
Block duration $\tau=M\times Tt_{\textrm{s}}$&$200\upmu$s\\ \hline
Train's velocity $v$&$0$ $\sim$ $200$m/s~\footnote{According to the system settings, coherence time is given by $t_{\textrm{c}}\approx \frac{0.423}{f_{D}}=\frac{4.23\times 10^{-2}}{v}$s. Recall the assumption that $Tt_{\textrm{s}}<t_{\textrm{c}}$, the train velocity must satisfy $v<211.5$m/s. Therefore, the range of $v$ is set to $0$ $\sim$ $200$m/s.}\\ \hline
Train's moving direction $\theta$&$0,~\frac{1}{16}\pi,~\frac{1}{8}\pi,~\frac{1}{4}\pi$\\ \hline
\end{tabular}
\end{center}
\end{minipage}
\end{table}

\begin{figure}[!t]\centering
\includegraphics[angle=0,scale=0.4]{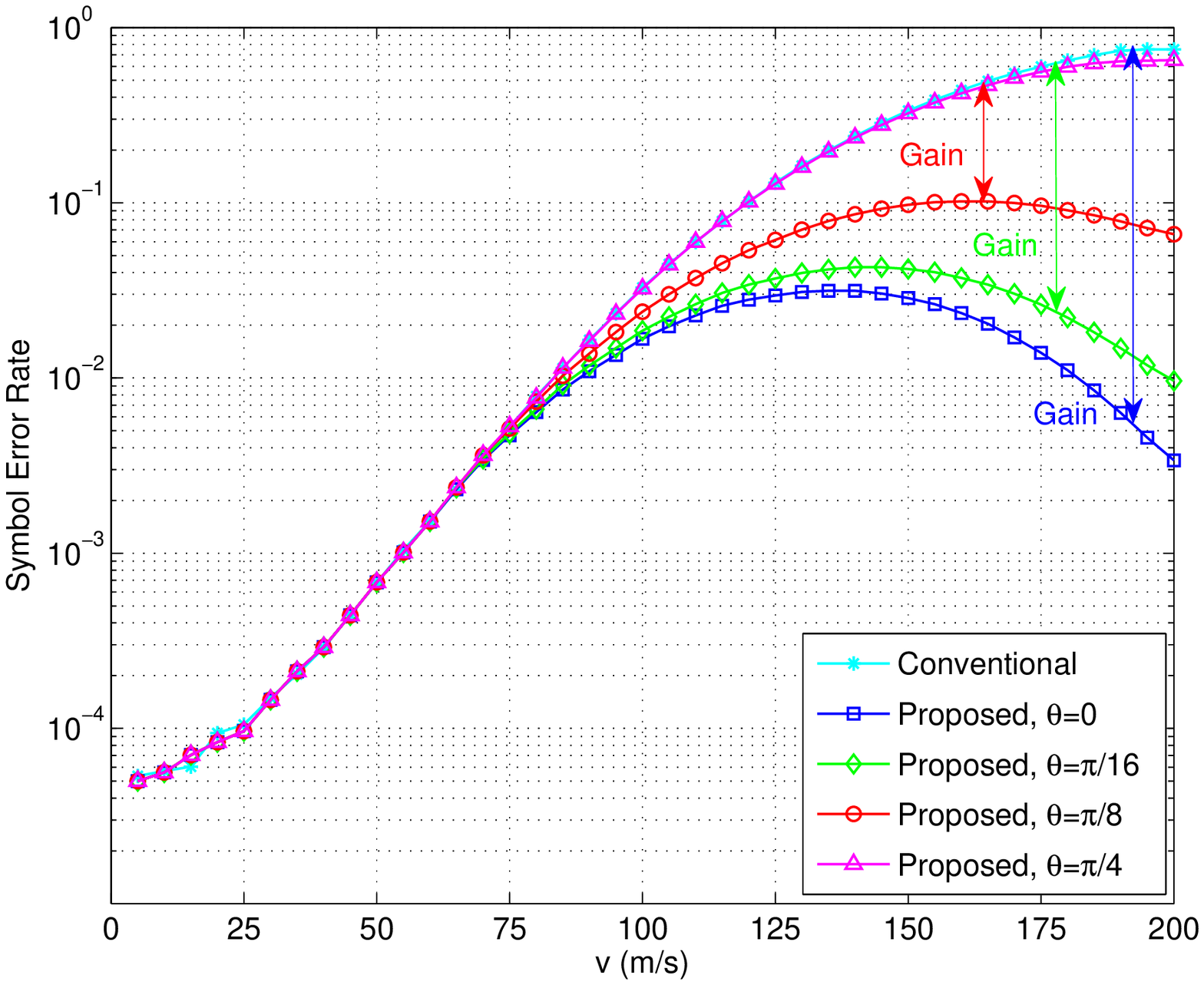}
\caption{Impacts of velocity and moving direction on the SER.}
\label{Fig_SERvsVelocity1}
\end{figure}

Fig. \ref{Fig_SERvsVelocity1} illustrates the SER of the proposed scheme with different velocities and moving directions. The conventional receiver (\ref{Standard-Decoding}) is employed as a benchmark. It is observed that, at the low mobility scenario (i.e., $v<75 \mbox{m/s}$), the conventional receiver achieves almost the same error performance as the proposed scheme. This is intuitive, since in low $v$ regime, the correlation matrix $\textbf{C}$ can be approximated as $\rho_{l}\textbf{I}_{N_{\textrm{R}}}$, indicating that the improved decoder is equivalent to the conventional one. Nevertheless, as the velocity increases, the error performance of the conventional receiver deteriorates substantially. In contrast, if the velocity is sufficiently large (i.e., $v>135 \mbox{m/s}$), the error performance of the proposed scheme actually improves as the velocity increases. This phenomenon will be elaborated in Section III.C. In addition, we see that the proposed scheme heavily depends on the moving direction. When $\theta > \frac{1}{4}\pi$, the performance gain becomes negligible. The reason is that the relative position has changed significantly, and the channels become almost uncorrelated, hence the proposed scheme only yields negligible performance gain.

\subsection{A Special Case}
\begin{figure}[h!t]
\centering
\subfigure[$\rho_{0},~\left| v\tau \right| < D-0.5\lambda$]{
\label{Fig-matched-velocity-low} 
\includegraphics[angle=0,scale=0.08]{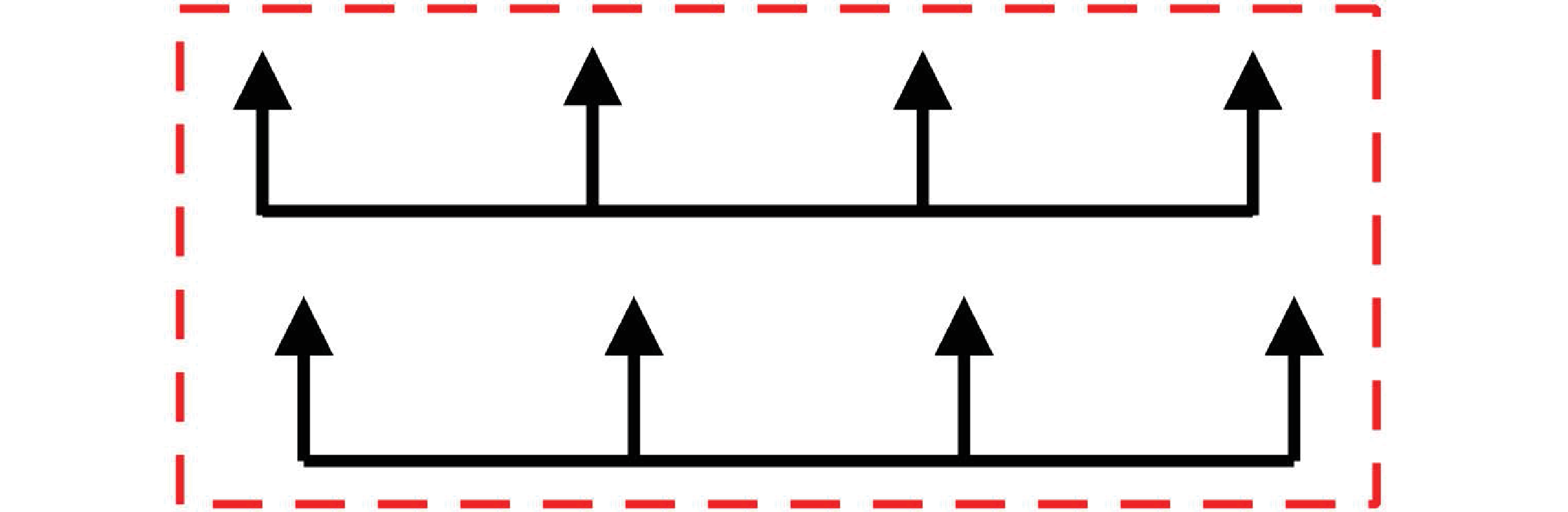}}
\hspace{0.24in}
\subfigure[$\rho_{1},~\left| v\tau-D \right| < D-0.5\lambda$]{
\label{Fig-matched-velocity-med} 
\includegraphics[angle=0,scale=0.08]{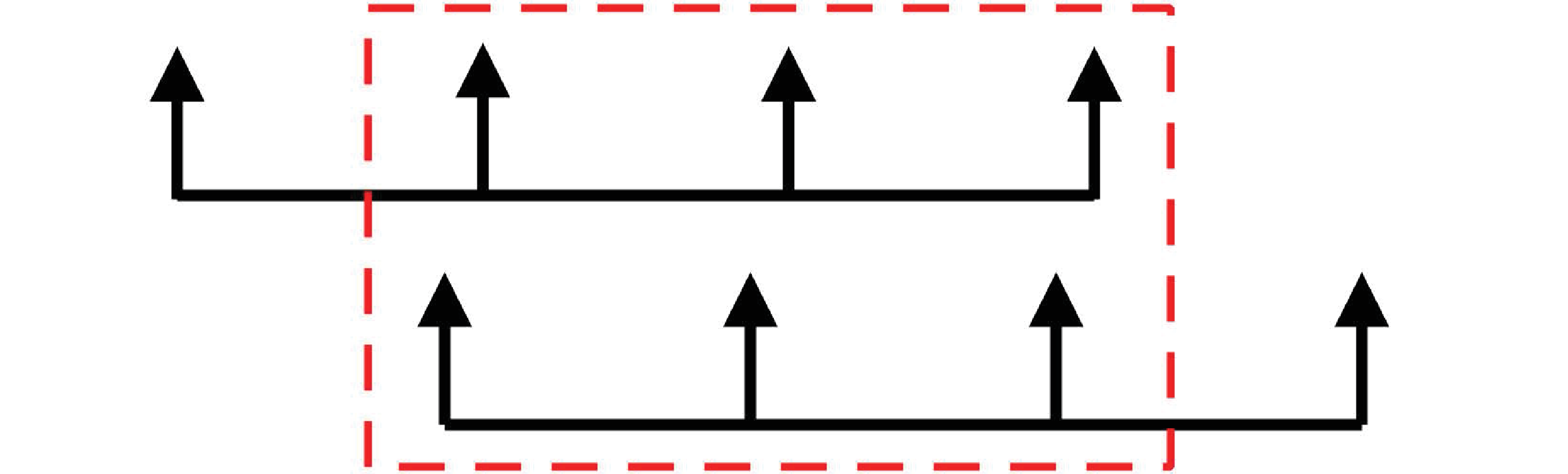}}
\hspace{0.12in}
\subfigure[$\rho_{2},~\left| v\tau-2D \right| < D-0.5\lambda$]{
\label{Fig-matched-velocity-high} 
\includegraphics[angle=0,scale=0.08]{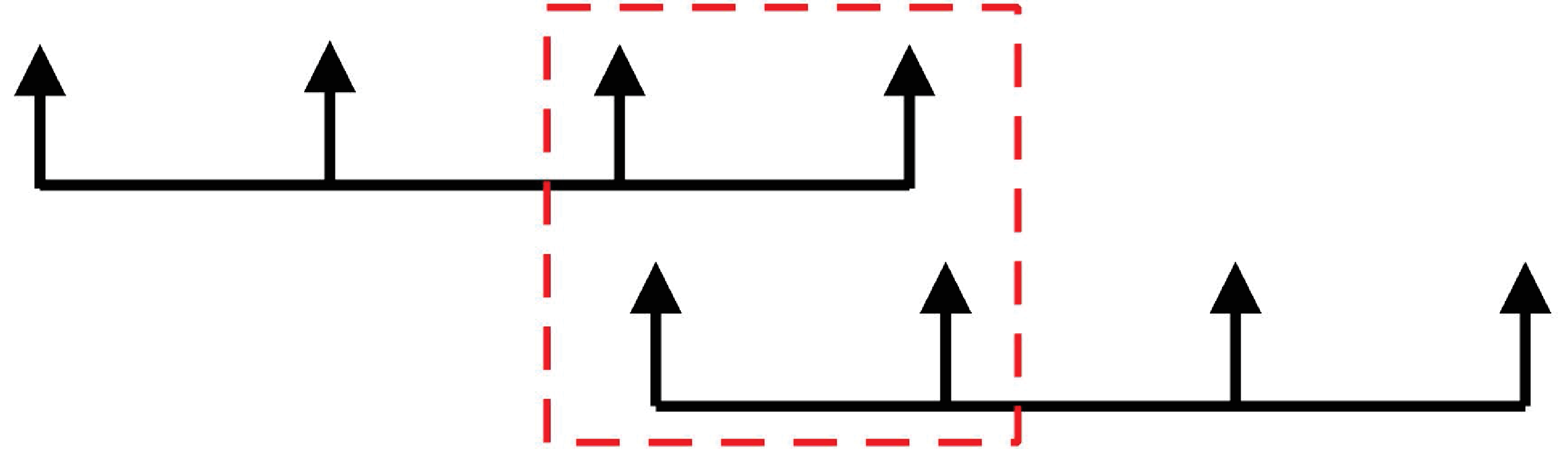}}
\hspace{0.12in}
\subfigure[$\rho_{3},~\left| v\tau-3D \right| < D-0.5\lambda$]{
\label{Fig-matched-velocity-uhigh} 
\includegraphics[angle=0,scale=0.08]{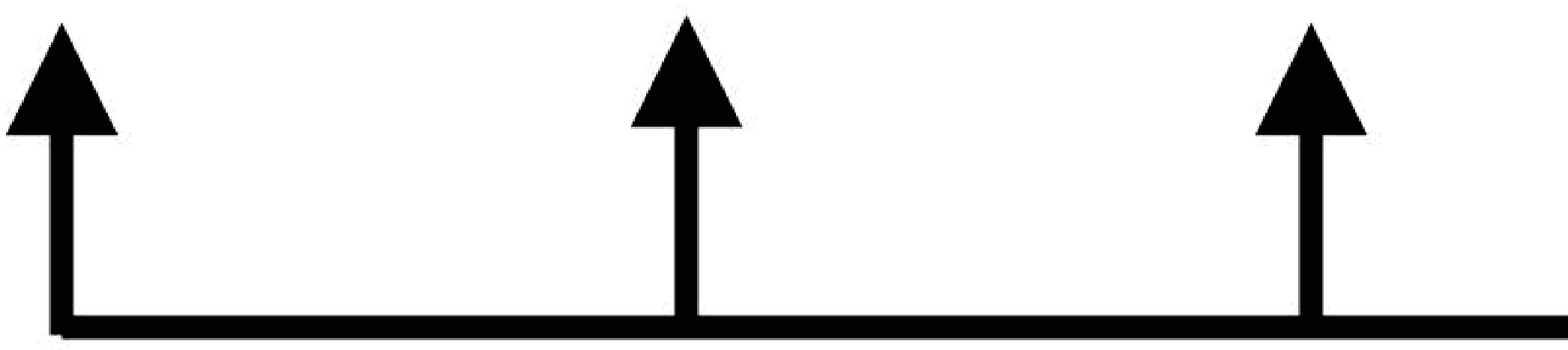}}
\caption{Impact of $v\tau$ on the spatial-temporal correlation structure.}
\label{Fig-matched-velocity} 
\end{figure}

In practice, the HST moves along the track with relatively constant velocity and the antenna array is always aligned with the direction of movement. Hence, in the ensuing analysis, we focus on the practical setting with a constant $v$ and $\theta=0$. In this sense, the channel correlation parameter given in (\ref{alpha_and_beta}) can be computed by
In practice, the HST moves along the track with relatively constant velocity and the antenna array is always aligned with the direction of movement. Hence, in the ensuing analysis, we focus on the practical setting with a constant $v$ and $\theta=0$. In this sense, the channel correlation parameter given in (\ref{alpha_and_beta}) can be computed by
\begin{equation}
\begin{split}
\rho_{l}&=J_{0}\left( \frac{2\pi}{\lambda}\left( v\tau - lD \right) \right)e^{-c_{0}v \tau }, \mbox{ and }\\
\rho_{l+1}=&
\left\{\begin{matrix}
J_{0}\left( \frac{2\pi}{\lambda}\left( v\tau - (l+1)D \right) \right)e^{-c_{0}v \tau },~\textrm{for case I}
\\
0,~~~~~~~~\textrm{for case II}~~~~~
\end{matrix}\right.
\label{alpha_and_beta2}
\end{split}
\end{equation}
When $ v\tau \approx {lD}$, i.e., the corresponding antenna pairs almost overlap at the same physical position at different time instant as illustrated in Fig. \ref{Fig-matched-velocity}, the correlation matrix ${\bf C} $ exhibits a special structure with only a single superdiagonal entry (i.e., case II). For such system, we have the following important result.

\begin{theorem}\label{theorem3}
For the special case where $\textbf{C}$ has one single superdiagonal (or main diagonal), the proposed decoding criterion (\ref{Improved-Decoding}) achieves the same error performance as the conventional differential decoder (\ref{Standard-Decoding}) with only $N_{\textrm{R}}-l$ receiving antennas.
\end{theorem}

\begin{IEEEproof}[Proof]
See Appendix \ref{proof_theorem3}.
\end{IEEEproof}

\begin{figure}[!t]\centering
\includegraphics[angle=0,scale=0.4]{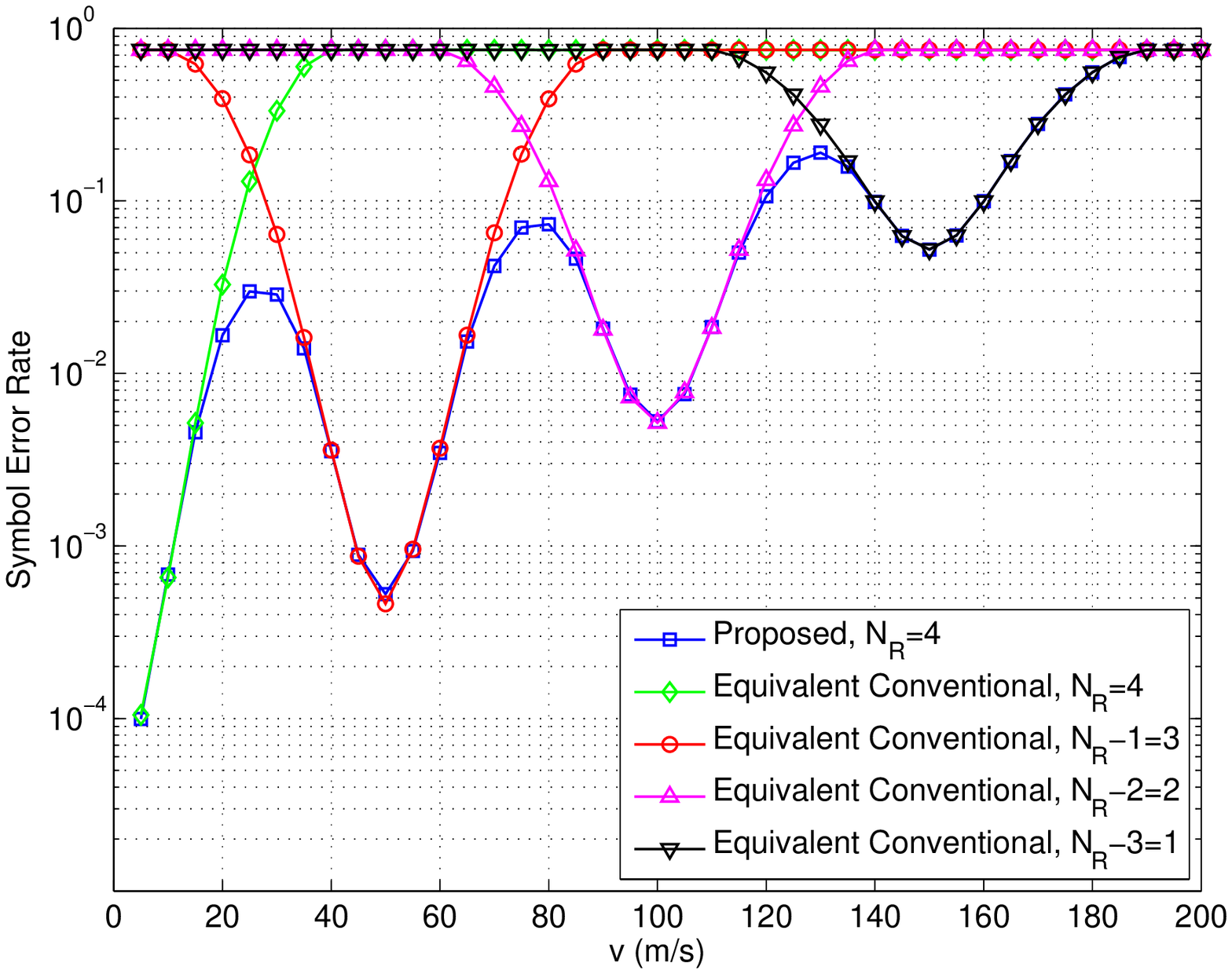}
\caption{SER induced by the proposed scheme versus the equivalent conventional one that exploits the signals received by antennas $\{ l+1,\dots,N_{\textrm{R}} \}$ at time $t_{i}$ and $\{ 1,\dots,N_{\textrm{R}}-l \}$ at time $t_{i+1}$ for differential decoding.}
\label{Fig_Theorem_2}
\end{figure}

\emph{Remarks}: Theorem \ref{theorem3} indicates that the penalty of the mobility is the requirement of extra receiving antennas at the relay. As illustrated in Fig. \ref{Fig_Theorem_2} where $M=5$, when $v\approx 0,~50,~100$ and $150$m/s, the proposed decoding criterion achieves the same error performance as an equivalent conventional receiver with only $N_{\textrm{R}}-l$ antennas.

Furthermore, a tighter PEP upper bound can be obtained as follows.
\begin{theorem}
When the correlation matrix ${\bf C}$ has one single superdiagonal (or main diagonal) with entries $\rho_{l}$, the PEP of the proposed differential scheme is upper bounded by
\begin{equation}
\begin{split}
&\textrm{Pr}\left \{ \textbf{G}\rightarrow {\textbf{G}}' \right \}\leq\\
&\frac{1}{\left | \textbf{I}_{N_\emph{T}}+\frac{\left( \frac{2\gamma_{1}\gamma_{2}}{\gamma_{1}+\gamma_{2}} \right)^{2}}{4N_{\emph{T}}\left(1+2\left( \frac{2\gamma_{1}\gamma_{2}}{\gamma_{1}+\gamma_{2}} \right) \right ) } \textbf{D}\left ( \textbf{G}-{\textbf{G}}' \right )\left ( \textbf{G}-{\textbf{G}}' \right )^{H}\textbf{D}^{H} \right |^{N_{\emph{R}}-l}},
\label{PEP-pro}
\end{split}
\end{equation}
where $\gamma_{1}$ and $\gamma_{2}$ are expressed as
\begin{equation}
\gamma_{1}=\frac{N_{\emph{T}}P}{\sigma^{2}_{\emph{n}}},~~~~~
\gamma_{2}=\frac{\left| \rho_{l} \right|^{2}}{1-\left| \rho_{l} \right|^{2}+\frac{\sigma^{2}_{\emph{n}}}{N_{\emph{T}}P}}.
\label{SINR-pro}
\end{equation}
\end{theorem}

\begin{IEEEproof}[Proof]
See Appendix \ref{proof_theorem4}.
\end{IEEEproof}

We now look into the high SNR regime. Recall from (20) that the performance of the proposed differential modulation scheme depends on the quality of two received signals. Because the second received symbol is subject to interference as shown in (19), the effective SINR settles as the transmit power increases, i.e., $\gamma_2 \rightarrow \frac{\left| \rho_{l} \right|^2}{1-\left| \rho_{l} \right|^2}$. Hence, we have
\begin{align}
\frac{2\gamma_{1}\gamma_{2}}{\gamma_{1}+\gamma_{2}} \approx 2\gamma_2\approx \frac{2\left| \rho_{l} \right|^2}{1-\left| \rho_{l} \right|^2}.
\end{align}
To this end, (\ref{PEP-pro}) reduces to
\begin{equation}
\begin{split}
&\textrm{Pr}\left \{ \textbf{G}\rightarrow {\textbf{G}}'|\textrm{SNR}\rightarrow \infty \right \}\leq\\
&\frac{1}{\left | \textbf{I}_{N_\emph{T}}+\frac{ \left| \rho_{l} \right|^{4}}{N_{\emph{T}}\left( 1+3\left| \rho_{l} \right|^{2} \right)\left( 1-\left| \rho_{l} \right|^{2} \right)  } \textbf{D}\left ( \textbf{G}-{\textbf{G}}' \right )\left ( \textbf{G}-{\textbf{G}}' \right )^{H}\textbf{D}^{H} \right |^{N_{\emph{R}}-l}},
\label{PEP-pro-highSNR}
\end{split}
\end{equation}
which indicates the existence of an error floor in the high SNR regime.

\begin{figure}[!t]\centering
\includegraphics[angle=0,scale=0.4]{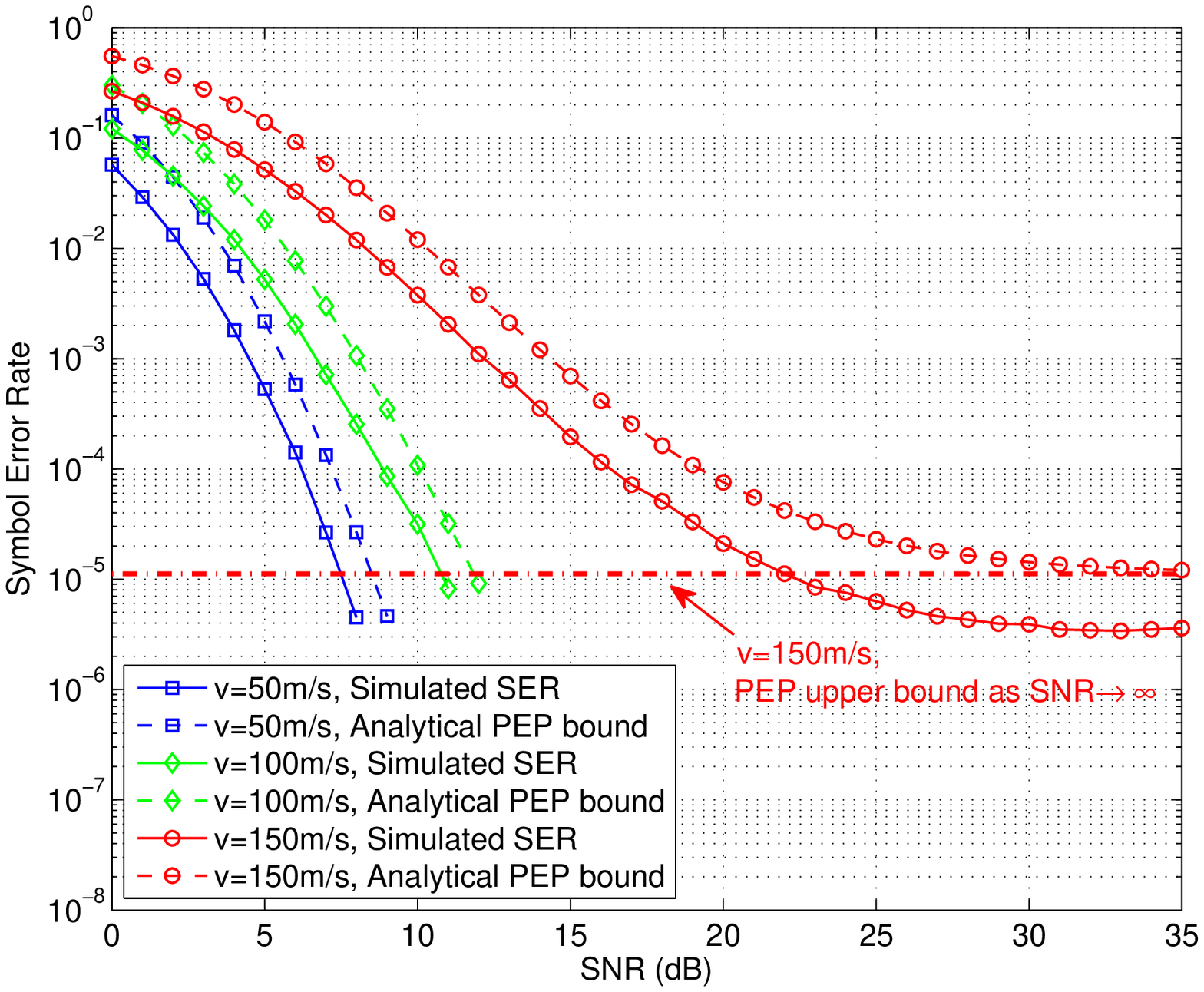}
\caption{Impact of SNR on SER and PEP upper bound when $v=50,100$ and $150$m/s.}
\label{Fig_SERvsSNR}
\end{figure}

Fig. \ref{Fig_SERvsSNR} illustrates the SERs and PEP upper bounds under different SNR conditions when $v=\frac{mD}{\tau},~m=1,\dots,N_{\textrm{R}}-1$. It is intuitive that lower velocity and higher SNR lead to better error performance. As expected, an error floor is observed as the SNR becomes large. The red dashed line represents the PEP upper bound when $v=150$m/s as $\textrm{SNR}\rightarrow \infty$.

\section{Enhanced Scheme with Transmit Adaptation}
In Section III, we have analyzed the error performance of the proposed scheme with fixed $v$ and $M$. In this section, we show that by adaptively choosing the block length $M$ according to the velocity of the train, substantial performance gain can be obtained. To see this, we first take a look at the impact of block length on the error performance.

\subsection{Impact of Block Length}
According to (\ref{Correlation_case_a}), (\ref{Correlation_case_b}) and (\ref{alpha_and_beta2}), for a fixed $D$, we have
\begin{equation}
l=\left\{\begin{matrix}
\lfloor \frac{vMTt_{\textrm{s}}}{D} \rfloor,~~~~~~~~\textrm{for case I}
\\
\mathrm{Round}\left\{ \frac{vMTt_{\textrm{s}}}{D} \right\},~~\textrm{for case II}~~~
\end{matrix}\right.
\end{equation}
\begin{equation}
\rho_{l}=J_{0}\left( \frac{2\pi}{\lambda}\left( vMTt_{\textrm{s}} - lD \right) \right)e^{-c_{0}vMTt_{\textrm{s}} },
\end{equation}
and
\begin{equation}
\begin{split}
\rho_{l+1}=&
\left\{\begin{matrix}
J_{0}\left( \frac{2\pi}{\lambda}\left( vMTt_{\textrm{s}} - (l+1)D \right) \right)e^{-c_{0}vMTt_{\textrm{s}} },\\
\textrm{for case I}
\\
0,~~~~~~~~~~~~~~~~~~\textrm{for case II}
\end{matrix}\right.
\label{alpha_and_beta_and location}
\end{split}
\end{equation}
It can be observed that $M$ has a great impact on the values of $l$, $\rho_{l}$ and $\rho_{l+1}$. Meanwhile, an insightful observation is that, as the velocity $v$ changes, if the transmitter can adjust $M$ adaptively such that $vM$ stays approximately the same, then the channel correlation statistics $\textbf{C}$ shall not change significantly. Recall the PEP bounds given in (\ref{PEP}) and (\ref{PEP-pro}), we see that if the channel correlation pattern $\textbf{C}$ remains unchanged, then the corresponding error performance also stays the same. Therefore, adaptation of the block length $M$ leads to a much robust PEP and SER performance, which sheds light on how to combat the velocity variation.

\begin{figure}[!t]\centering
\includegraphics[angle=0,scale=0.4]{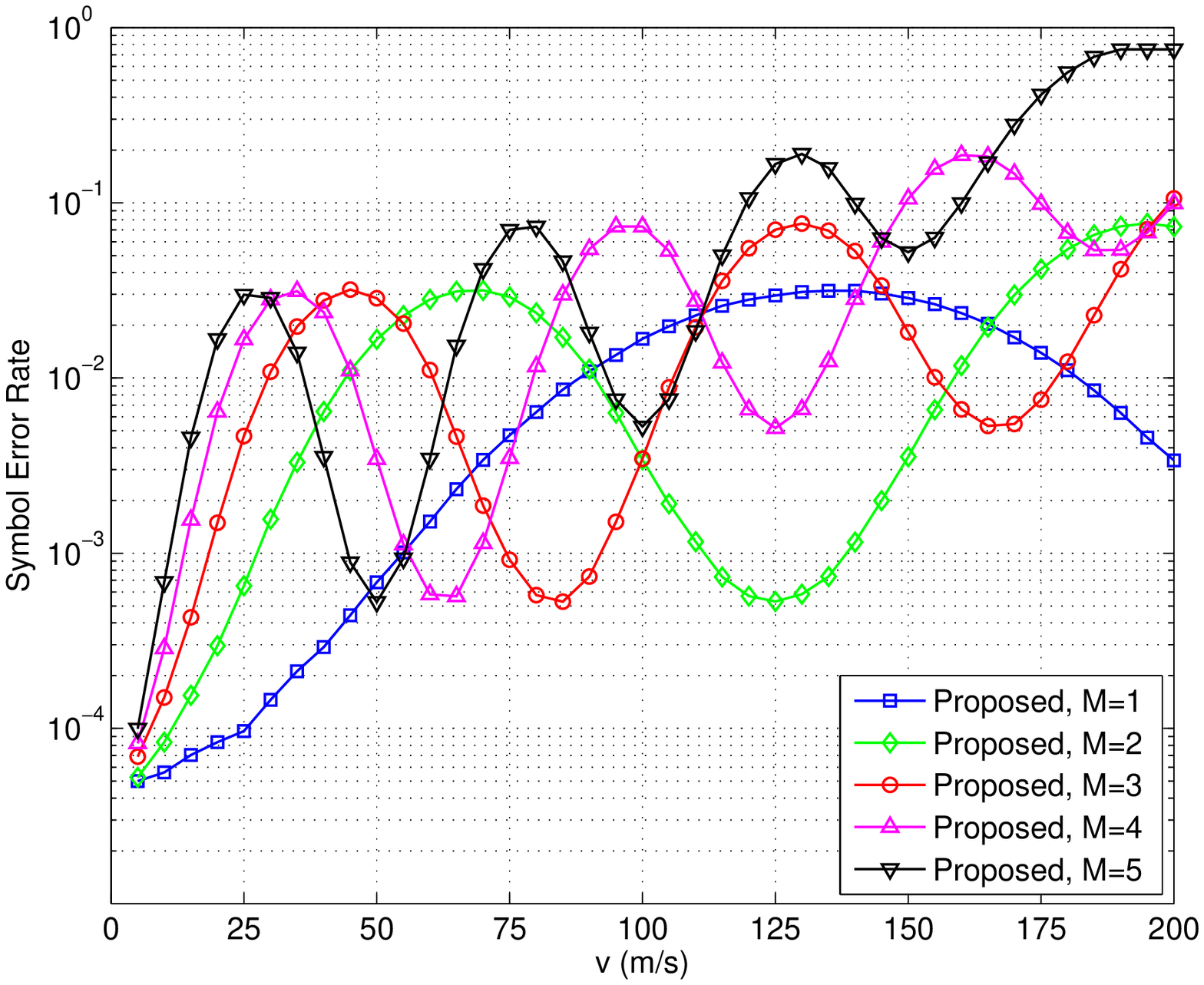}
\caption{Impact of block length $M$ on the SER performance.}
\label{Fig_SERvsTransInterval}
\end{figure}

We now illustrate the impact of block length on system error performance through the following example. For the simulation, we choose $M=1$, $2$, $3$, $4$, $5$ while other parameters remain the same as in Section III. From Fig. \ref{Fig_SERvsTransInterval}, we see that different block lengths result in distinct SER curves. Moreover, significant fluctuation in SER is observed when $M$ gets larger. This is rather intuitive, since a larger $M$ amplifies the influence of velocity variation on the channel correlation structure $\textbf{C}$, hence causing more frequent fluctuations. Furthermore, when $v=50$, $65$, $85$ and $125$m/s, by properly choosing $M$, i.e., $M = 5, 4, 3, 2$, the achievable SER remains the same, which corroborates the finding of Theorem \ref{theorem3}.

\subsection{Adaptive Transmission Method}
We now address the problem of how to adaptively set the block length $M$ as the velocity $v$ changes. Since the motivation of adaptation is to improve the error performance, we consider the optimal $M$ being the solution of the following optimization problem:
\begin{align*}
&\min_{M}\quad \textrm{PEP}_{\textrm{bound}}\left(v,M \right)\\
&~s.t.\quad M\in \mathbb{Z^{+}}
\end{align*}
where $\textrm{PEP}_{\textrm{bound}}$ follows (\ref{PEP}) for case I and (\ref{PEP-pro}) for case II. Since $M$ is a positive integer, we can obtain the optimal solution through exhaustive search (for clarity purposes, we denote this method as OptM). However, since there is no upper limit to the value of $M$, OptM may cause an intolerable overhead. To tackle this problem, we propose a much easier heuristic adaptation method, conserving computing resources while providing robustness.

Before presenting our method, let us revisit the numerical results in Fig. \ref{Fig_SERvsTransInterval}. When the train velocity is low (i.e., $v<50$m/s), we see $M=1$ provides the best SER performance. This is intuitive, since the successive channel samples are strongly correlated in time domain in this case. However, when the train velocity exceeds a certain threshold $v_{0}$ (i.e., $50$m/s), unit block length incurs a noticeable performance penalty due to the fact that channel's variation becomes increasingly significant. To address this issue, an intuitive solution is that the BS dynamically adjusts $M$ so that $vMTt_{\textrm{s}}\approx mD,~m=0,\dots,N_{\textrm{R}}-1$ (as depicted in Fig. \ref{Fig-matched-velocity}). As such, the influence of velocity variation is suppressed and, in the meantime, the level of spatial-temporal channel correlation will always be high enough to support differential decoding. As it is impossible to satisfy this criteria when $m=0$, we consider the case $m=1$, i.e., $vMTt_{\textrm{s}}\approx D$.

Now we are ready to present our heuristic transmit adaptation (HTA) method:

\emph{a)} In low $v$ regime (i.e., $v<v_{0}$), the BS exploits the proposed scheme with $M=1$.

\emph{b)} When $v$ exceeds the threshold $v_{0}$, the BS adaptively chooses $M$ according to $v$, so that $vMTt_{\textrm{s}}\approx D$. Given that $M\geq 1$, we have $M=\max \left\{ \mathrm{Round}\left\{ \frac{D}{vTt_{\textrm{s}}} \right\},1\right\}$.

\begin{figure}[!t]
\centering
\subfigure[Symbol error rates affected by different TA schemes.]{
\label{Fig-TA} 
\includegraphics[angle=0,scale=0.4]{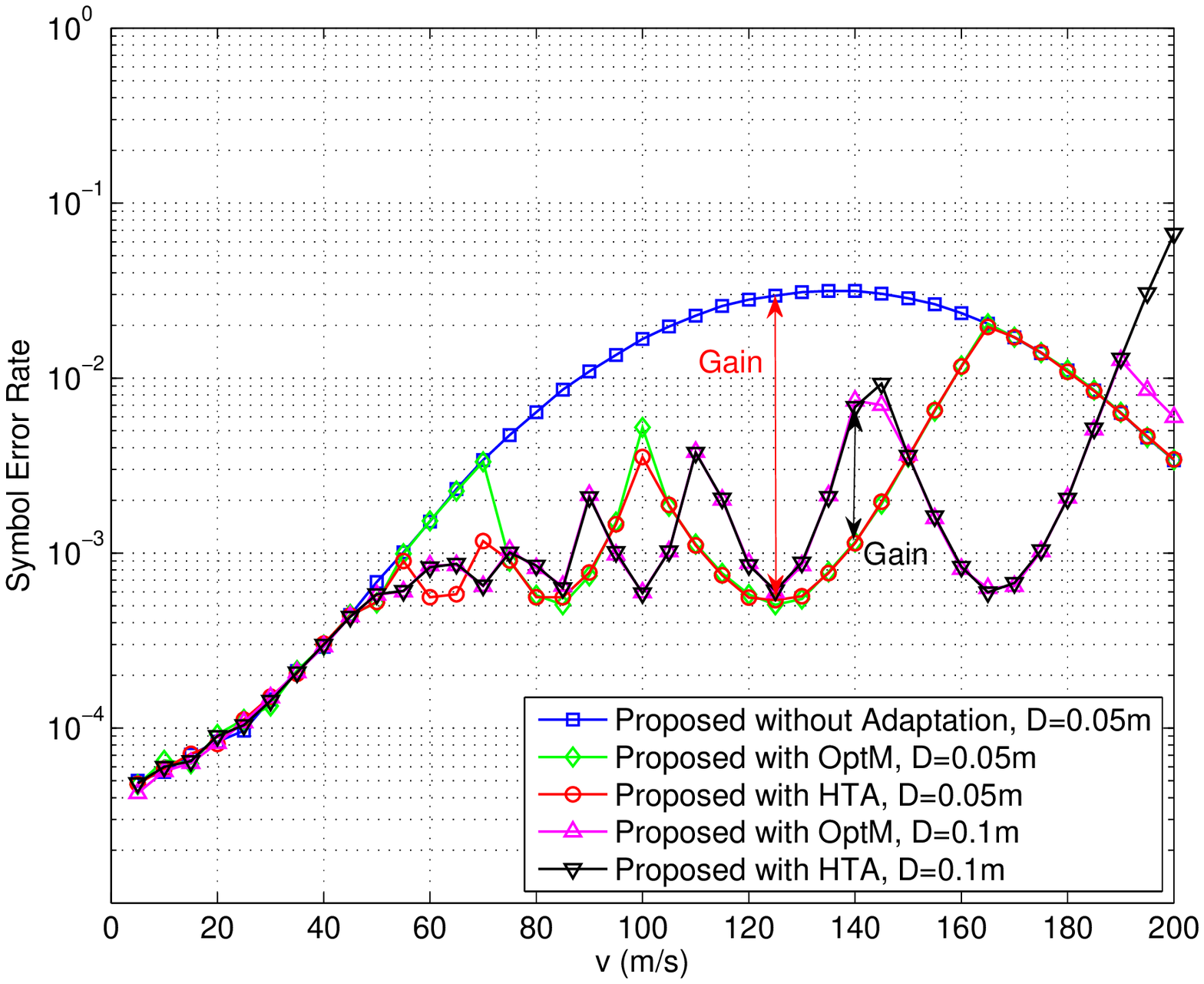}}
\hspace{0.4in}
\subfigure[The discrete values of $M$ chosen by the BS.]{
\label{Fig-M} 
\includegraphics[angle=0,scale=0.4]{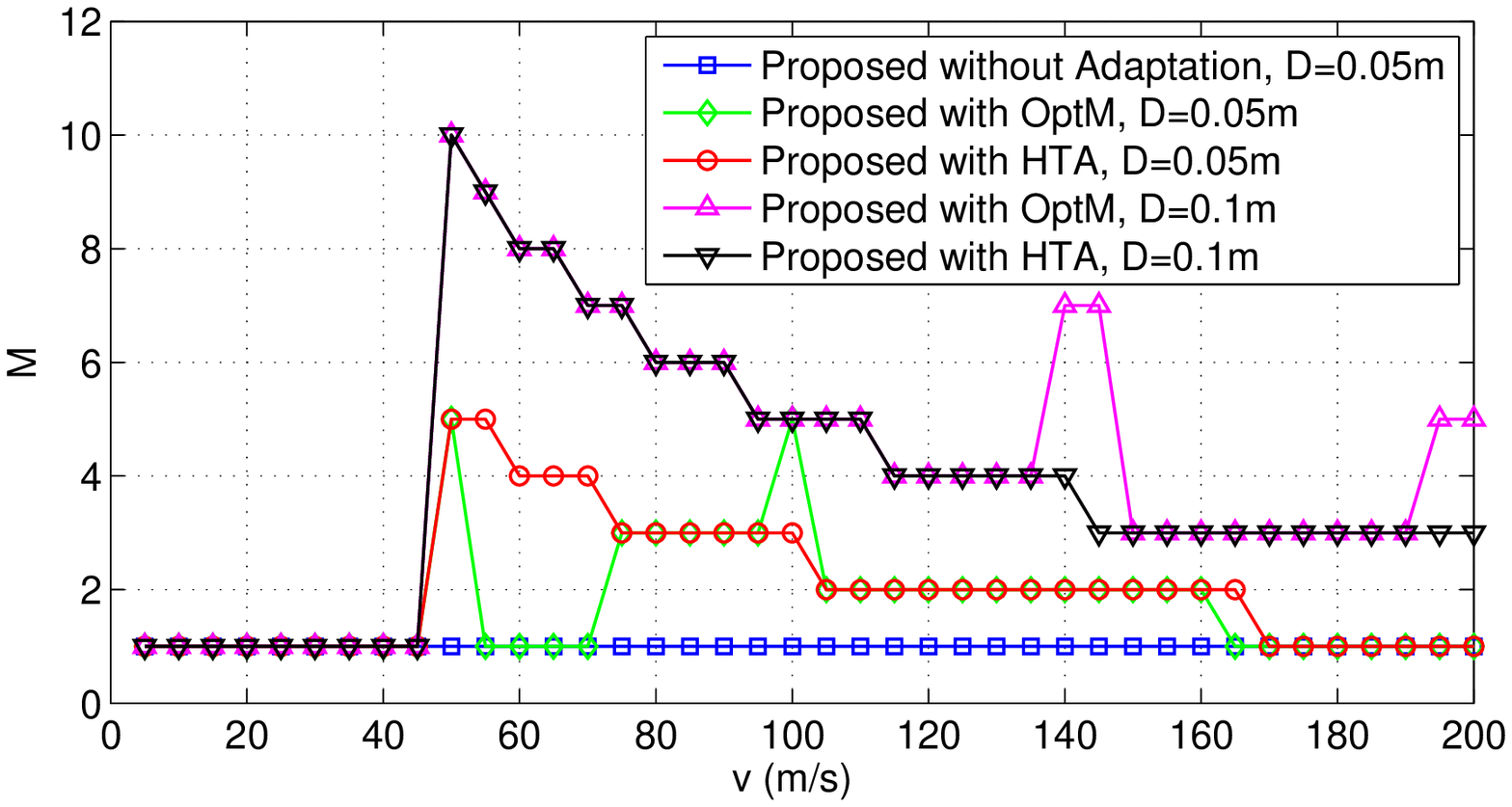}}
\caption{SER performance and the corresponding $M$ induced by different adaptation methods with distinct antenna spacings.}
\label{Fig-TA-and-M} 
\end{figure}

Now the remaining task lies in the threshold $v_{0}$. This is rather sample, since $v_{0}$ is the minimum solution of the following equation
\begin{equation}
\textrm{PEP}_{\textrm{bound}}\left( v,1\right)=\textrm{PEP}_{\textrm{bound}}\left( v,\mathrm{Round}\left\{ \frac{D}{vTt_{\textrm{s}}} \right\} \right).
\label{Velocity-Threshold}
\end{equation}

\newcounter{TempEqCnt3}
\setcounter{TempEqCnt3}{\value{equation}}
\setcounter{equation}{41}
\begin{figure*}[hb]
\hrulefill
\begin{equation}
\begin{split}
\widehat{\textbf{G}}
~=~&\mathrm{arg}\underset{\textbf{G}\in \mathcal{G}}{\mathrm{min}}\left \| \textbf{Y}_{i+1}-\textbf{C}\textbf{Y}_{i}\textbf{G} \right \|_{\textrm{F}}\\
~=~&\mathrm{arg}\underset{\textbf{G}\in \mathcal{G}}{\mathrm{min}}\textrm{Tr}\left\{ \textbf{Y}_{i+1}\textbf{Y}_{i+1}^{H}-\textbf{Y}_{i+1}\textbf{G}^{H}\textbf{Y}_{i}^{H}\textbf{C}^{H}-\textbf{C}\textbf{Y}_{i}\textbf{G}\textbf{Y}_{i+1}^{H}+\textbf{C}\textbf{Y}_{i}\textbf{G}\textbf{G}^{H}\textbf{Y}_{i}^{H}\textbf{C}^{H} \right\}\\
~=~&\mathrm{arg}\underset{\textbf{G}\in \mathcal{G}}{\mathrm{min}}\textrm{Tr}\left\{ \textbf{Y}_{i+1}\textbf{Y}_{i+1}^{H}-\textbf{Y}_{i+1}\textbf{G}^{H}\textbf{Y}_{i}^{H}\textbf{C}^{H}-\textbf{C}\textbf{Y}_{i}\textbf{G}\textbf{Y}_{i+1}^{H}+\textbf{C}\textbf{Y}_{i}\textbf{Y}_{i}^{H}\textbf{C}^{H} \right\}\\
~=~&\mathrm{arg}\underset{\textbf{G}\in \mathcal{G}}{\mathrm{max}}\textrm{Re}\left\{ \textrm{Tr} \left\{  \textbf{C}\textbf{Y}_{i}\textbf{G}\textbf{Y}_{i+1}^{H}   \right\}  \right\}
\label{Euclidean-detector}
\end{split}
\end{equation}
\end{figure*}
\setcounter{equation}{\value{TempEqCnt3}}

In Fig. \ref{Fig-TA-and-M}(a), we show numerically the error performance induced by different adaptation methods with distinct antenna spacings. For the case $D=0.05$m, invoking the non-adaptive scheme as the benchmark, we see that both OptM and HTA enhance the system robustness. It should be noted that the exhaustive-search based OptM scheme intends to minimize the PEP bound, not the real PEP or SER, hence it is possible that the OptM scheme performs worse than HTA for some particular choice of D = 0.05m. Yet, as shown in Fig. 10(a), except these particular cases, the OptM and HTA achieve almost the same SER performance. On this basis, it is reasonable to treat HTA as a near-optimal solution to the above optimization problem. Considering its low computational complexity, HTA may be a preferred option in high mobility settings. Moreover, for the proposed HTA scheme, it can be observed that setting $D=0.05$m outperforms the case $D=0.1$m when $v=110\sim 150$m/s. This observation indicates that, to fully exploit the spatial-temporal correlation, antenna spacing should be carefully designed if the statistical information (i.e., mean and variance) of the train velocity is known to the BS. Finally, Fig. \ref{Fig-TA-and-M}(b) depicts the corresponding values of $M$. When $D=0.05$m and $0.1$m, velocity thresholds $v_{0}$ are about $47$m/s and $48$m/s, respectively. Thereby, block lengths are both set to be $1$ when $v=0\sim 45$m/s. As $v$ exceeds $v_{0}$, a substantial increase in $M$ is observed. After that, $M$ will decrease gradually since $vMTt_s$ needs to be aligned to $D$. When it returns to $1$, HTA becomes of little avail because the train velocity is too high.

\section{Conclusion}
We have studied the joint spatial and temporal channel correlation inherent in high mobility scenarios with a moving antenna array. Capitalizing on this type of correlation information, an improved DSTM scheme was proposed, and the corresponding error performance was analyzed through the characterization of the PEP upper bound. Finally, an adaptive transmission scheme varying the block length according to the node velocity was proposed. The findings of the paper suggest that the proposed DSTM scheme outperforms the conventional scheme, and further performance improvement can be realized by adaptively adjusting the block length.

\appendix
\subsection{Proof of Lemma 1}\label{proof_lemma1}
\begin{figure}[ht]\centering
\includegraphics[angle=0,scale=0.15]{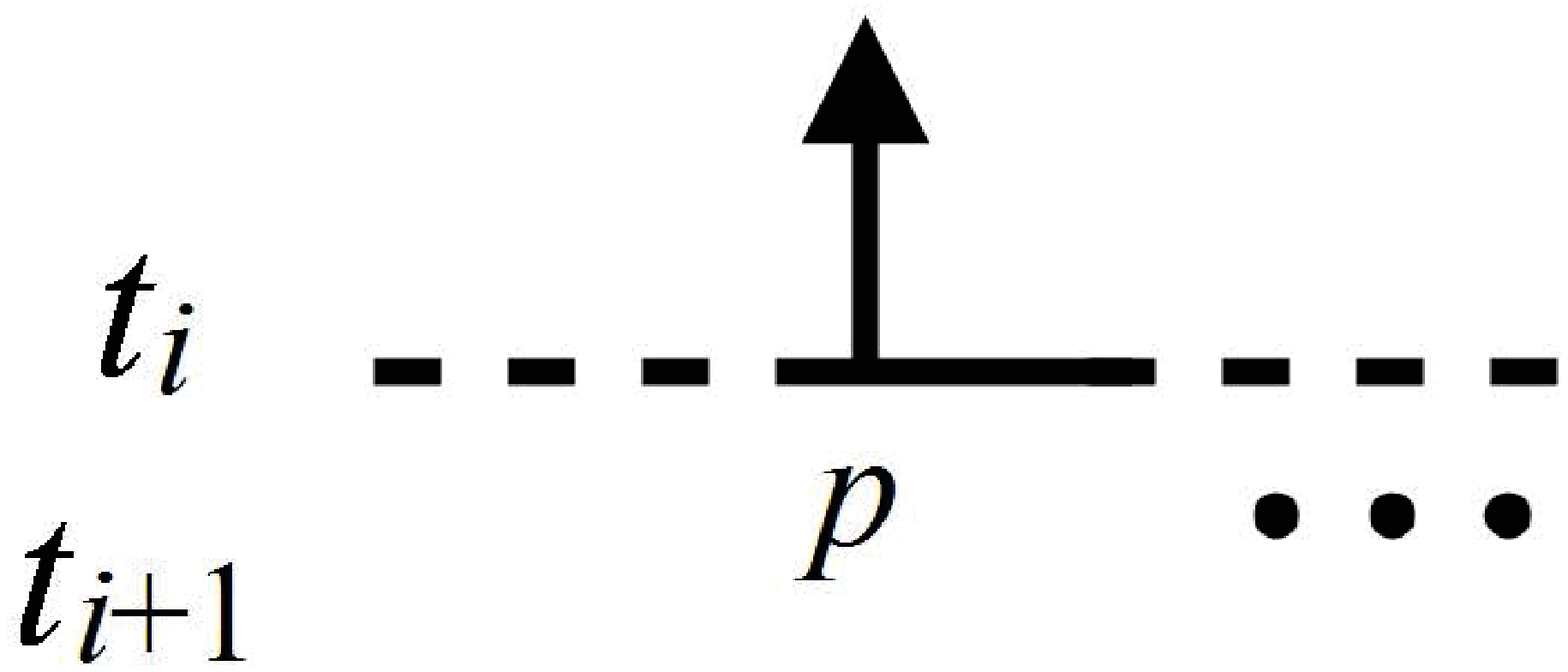}
\caption{The channels between the $m$-th BS antenna and the moving antenna array.}
\label{Fig_Correlation_Model}
\end{figure}

Fig. \ref{Fig_Correlation_Model} illustrates the channel changes between the $m$-th BS antenna and the moving antenna array at time $t_{i}$ and $t_{i+1}$. Denote $h_{m,n,i}$ as the channel gain corresponding to the $n$-th HST antenna at time $t_{i}$.

Given adequately large distance between the BS and HST, $\mu$ (the mean direction of AOA) can be considered the same for different antennas. Substituting this condition into (\ref{modi_correlation_model}), it can be inferred that the statistical correlation between $h_{m,p,i+1}$ and $h_{m,q,i}$ is approximately equal to that between $h_{m,p+1,i+1}$ and $h_{m,q+1,i}$, i.e., for $1\leq p,q \leq N_{\textrm{R}}-1$,
\begin{equation}
\textrm{E}[h_{m,p,i+1}h_{m,q,i}^{\ast}]=\textrm{E}[h_{m,p+1,i+1}h_{m,q+1,i}^{\ast}].
\label{Toeplitz-correlation-proof1}
\end{equation}

Recall that $h_{m,n,i}$ is the $n$-th entry of $\textbf{h}_{m,i}$, it can be inferred from (\ref{Correlation-Model}) that
\begin{equation}
[\textbf{C}]_{p,q}=\textrm{E}[h_{m,p,i+1}h_{m,q,i}^{\ast}].
\label{Toeplitz-correlation-proof2}
\end{equation}

Combining (\ref{Toeplitz-correlation-proof1}) and (\ref{Toeplitz-correlation-proof2}), the following equation holds:
\begin{equation}
[\textbf{C}]_{p,q}=[\textbf{C}]_{p+1,q+1},
\label{Toeplitz-correlation-proof}
\end{equation}
which indicates that the correlation matrix $\textbf{C}$ is a Toeplitz matrix.

Now we divide the remaining proof into the following two cases:

\subsubsection{The receive array is motionless, i.e., $v=0$ and $\theta=0$}
In this case, the scattering environment remains the same, and therefore, the channel exhibits static property. As a consequence, the correlation matrix is $\textbf{I}_{N_{\textrm{R}}}$, which coincides with (\ref{Toeplitz-correlation-Model}).

\subsubsection{The receiver is moving}
Consider the scenario in Fig. \ref{Fig_Correlation_Model}. Since the $q$-th and $(q+1)$-th receive antenna at time $t_{i}$ are probably within the coherent distance of the $p$-th receive antenna at time $t_{i+1}$, they share similar scattering environments. On this basis, $h_{m,p,i+1}$ is most interrelated to $h_{m,q,i}$ and $h_{m,q+1,i}$, with correlation coefficients $\rho_{q-p}$ and $\rho_{q-p+1}$, respectively. As for other antennas, Fact 1 states that their channel gains are all uncorrelated with $h_{m,p,i+1}$. Therefore, the correlation matrix is a Topelitz matrix with $\rho_{q-p}$ and $\rho_{q-p+1}$ on its $(q-p)$-th and $(q-p+1)$-th superdiagonal. Replacing $q-p$ with $l$ completes the proof.

\subsection{Proof of Theorem 1}\label{proof_theorem1}
From (\ref{Two-Received}), it is easy to obtain:
\begin{equation}
\textbf{Y}_{i+1}-\textbf{C}\textbf{Y}_{i}\textbf{G}_{k}=-\textbf{C}\textbf{N}_{i}\textbf{G}_{k}+\sqrt{P}\textbf{U}_{i+1}\textbf{X}_{i}\textbf{G}_{k}+\textbf{N}_{i+1}.
\label{Two-Received-pro}
\end{equation}
Since the statistics of $\textbf{N}_{i}$ and $\textbf{U}_{i+1}$ remain unchanged when multiplied with unitary matrices, we may rewrite (\ref{Two-Received-pro}) as
\begin{equation}
\textbf{Y}_{i+1}-\textbf{C}\textbf{Y}_{i}\textbf{G}_{k}=-\textbf{C}\textbf{N}_{i}'+\sqrt{N_{\textrm{T}}P}\textbf{U}_{i+1}'+\textbf{N}_{i+1}\triangleq \textbf{W}.
\label{Two-Received-pro2}
\end{equation}
Due to that the columns of $\textbf{N}_{i}'$, $\textbf{U}_{i+1}'$ and $\textbf{N}_{i+1}$ are mutually independent complex Gaussian random vectors, each column of $\textbf{W}$ is also complex Gaussian with a covariance matrix
\begin{equation}
\begin{split}
\textbf{R}_{\textrm{w}}
&=\left(-\textbf{C}\right)\textbf{R}_{\textrm{n}}\left(-\textbf{C}\right)^{H} + \sqrt{N_{\textrm{T}}P}\textbf{R}_{\textrm{u}}\sqrt{N_{\textrm{T}}P} + \textbf{R}_{\textrm{n}}\\
&=\sigma_{\textrm{n}}^{2}\textbf{C}\textbf{C}^{H} + N_{\textrm{T}}P\left( \textbf{I}_{N_{\textrm{R}}}-\textbf{C}\textbf{C}^{H} \right) + \sigma_{\textrm{n}}^{2}\textbf{I}_{N_{\textrm{R}}}.
\label{W-covariance}
\end{split}
\end{equation}
Therefore, the optimal ML receiver is as shown in (\ref{Euclidean-detector}). Combining with the fact $\textrm{Tr}\left\{ \textbf{AB}\right\}=\textrm{Tr}\left\{ \textbf{BA}\right\}$, the proof is complete.

\setcounter{equation}{42}

\newcounter{TempEqCnt4}
\setcounter{TempEqCnt4}{\value{equation}}
\setcounter{equation}{47}
\begin{figure*}[ht]
\begin{equation}
\gamma
=\frac{\textrm{E}\left [ \left \| \sqrt{P} \textbf{C}\textbf{H}_{i}\textbf{X}_{i}\textbf{G} \right \|_{F}^{2} \right ]}
{\textrm{E}\left [ \left \| \sqrt{P}\textbf{U}_{i+1}\textbf{X}_{i}\textbf{G}+\textbf{N}_{i+1} \right \|_{F}^{2} \right ]}
=\frac{N_{\textrm{T}}^{2}P\left [ \left ( N_{\textrm{R}}-l \right )\left | \rho_{l} \right |^{2}+\left ( N_{\textrm{R}}-l-1 \right )\left | \rho_{l+1} \right |^{2} \right ]}
{N_{\textrm{T}}^{2}P\left [ N_{\textrm{R}}- \left ( N_{\textrm{R}}-l \right )\left | \rho_{l} \right |^{2}-\left ( N_{\textrm{R}}-l-1 \right )\left | \rho_{l+1} \right |^{2} \right ]+N_{\textrm{R}}N_{\textrm{T}}\sigma^{2}_{\textrm{n}} },
\label{gamma}
\end{equation}
\hrulefill
\end{figure*}
\setcounter{equation}{\value{TempEqCnt4}}

\subsection{Proof of Theorem 2}\label{proof_theorem2}

Recall the Chernoff bound of the PEP of the conventional receiver for differential unitary space-time codes in quasi-static channels given by (cf.\cite{hughes2000differential}, eq.(9))
\begin{equation}
\textrm{Pr}\left\{ \bar{\textbf{C}}_{\textbf{G}}\rightarrow \bar{\textbf{C}}_{{\textbf{G}}'} \right\}\leq
\frac{1}{\left| \textbf{I}+\frac{\left(\frac{\rho}{t} \right)^{2}\bar{n}^{2}}{4\left( 1+\left(\frac{\rho}{t} \right)\bar{n} \right) }\left[ \textbf{I}-\frac{1}{ \bar{n}^{2}}\bar{\textbf{C}}_{\textbf{G}}\bar{\textbf{C}}_{{\textbf{G}}'}^{H}\bar{\textbf{C}}_{{\textbf{G}}'}\bar{\textbf{C}}_{\textbf{G}}^{H} \right] \right|^{r}},
\label{Original-PEP}
\end{equation}
where $\bar{n}$ is the length of a unitary code, $t$ and $r$ are the numbers of transmit and receive antennas, $\rho$ is the signal-to-noise ratio (SNR) per receive antenna, $\bar{\textbf{C}}_{\textbf{G}}$ and $\bar{\textbf{C}}_{{\textbf{G}}'}$ are the codeword matrices associated with the information matrices $\textbf{G}$ and ${\textbf{G}}'$. By associating our proposed differential scheme with the conventional scheme, we can then get the PEP performance.

To this end, first, we define $\bar{\bar{\textbf{X}}}_{i+1,\textbf{G}}\triangleq\left [ \textbf{X}_{i}~~ \textbf{X}_{i}\textbf{G} \right ]$ and $\bar{\bar{\textbf{Y}}}_{i+1}\triangleq\left [ \textbf{C}\textbf{Y}_{i}~~ \textbf{Y}_{i+1} \right ]$. Invoking the orthogonality properties (\ref{Orthogonal-Design}), we have
\begin{equation}
\bar{\bar{\textbf{X}}}_{i+1,\textbf{G}}\bar{\bar{\textbf{X}}}_{i+1,\textbf{G}}^{H}=2N_{\textrm{T}}\textbf{I}_{N_{\textrm{T}}}.
\label{Orthogonal-block}
\end{equation}
According to \cite{hughes2000differential}, $\bar{\bar{\textbf{X}}}_{i+1}$ can be treated as a unitary space-time code with length $2N_{\textrm{T}}$. This can be further revealed from the fact that, the optimal receiver defined in Theorem \ref{theorem1} can be rewritten as
\begin{equation}
\begin{split}
\widehat{\textbf{G}}
&=\mathrm{arg}\underset{\textbf{G}\in \mathcal{G}}{\mathrm{max}}\mathrm{Re}\left \{ \textrm{Tr} \left \{\textbf{G}\textbf{Y}_{i+1}^{H}\textbf{C}\textbf{Y}_{i} \right \}\right \}\\
&=\mathrm{arg}\underset{\textbf{G}\in \mathcal{G}}{\mathrm{max}}\textrm{Tr}\left\{\textbf{Y}_{i+1}\textbf{G}^{H}\textbf{Y}_{i}^{H}\textbf{C}^{H}+\textbf{C}\textbf{Y}_{i}\textbf{G}\textbf{Y}_{i+1}^{H} \right\}\\
&=\mathrm{arg}\underset{{\textbf{G}}\in \mathcal{G}}{\mathrm{max}}\textrm{Tr} \left \{ \bar{\bar{\textbf{Y}}}_{i+1}\bar{\bar{\textbf{X}}}_{i+1,\textbf{G}}^{H}\bar{\bar{\textbf{X}}}_{i+1,\textbf{G}}\bar{\bar{\textbf{Y}}}_{i+1}^{H}    \right \},
\end{split}
\label{Improved-Decoding2}
\end{equation}
resulting in a conventional \emph{quadratic receiver}. Note the last step of (\ref{Improved-Decoding2}) is due to that $\bar{\bar{\textbf{X}}}_{i+1,\textbf{G}}^{H}\bar{\bar{\textbf{X}}}_{i+1,\textbf{G}}=\small
{N_{\textrm{T}}} \begin{bmatrix}
\textbf{I}_{N_{\textrm{T}}} &\textbf{G} \\
\textbf{G}^{H} &\textbf{I}_{N_{\textrm{T}}}
\end{bmatrix}$. Therefore, the PEP performance of our proposed scheme is equivalent to that of the original unitary coding scheme with  $\bar{n}=2N_{\textrm{T}}$, $t=N_{\textrm{T}}$, $\bar{\textbf{C}}_{\textbf{G}}=\bar{\bar{\textbf{X}}}_{i+1,\textbf{G}}$ and $\bar{\textbf{C}}_{{\textbf{G}}'}=\bar{\bar{\textbf{X}}}_{i+1,{\textbf{G}}'}$.

Second, we proceed to find the equivalent $r$ and $\rho$, which may be a little tricky.

Recalling that (\ref{Improved-Decoding}) is based on maximizing the correlation between
\begin{equation}
\textbf{C}\textbf{Y}_{i}\textbf{G}=\sqrt{P}\textbf{C}\textbf{H}_{i}\textbf{X}_{i}\textbf{G}+\textbf{C}\textbf{N}_{i}\textbf{G}
\end{equation}
and
\begin{equation}
\textbf{Y}_{i+1}=\sqrt{P}\textbf{C}\textbf{H}_{i}\textbf{X}_{i}\textbf{G}+\sqrt{P}\textbf{U}_{i+1}\textbf{X}_{i}\textbf{G}+\textbf{N}_{i+1},
\label{Two-Received-pro3}
\end{equation}
we can thus regard $\textbf{C}\textbf{H}_{i}$ as the effective channel matrix. Keeping Lemma \ref{lemma1} in mind, we know that the entries in row $N_{\textrm{R}}-l+1 \sim N_{\textrm{R}}$ of $\textbf{C}\textbf{H}_{i}$ are all zeros. This corresponds to a receiver that comprises only $N_{\textrm{R}}-l$ antennas, leading to the fact that $r=N_{\textrm{R}}-l$.

On the other hand, from (\ref{Two-Received}), the average signal-to-interference-plus-noise ratios (SINR) after the second reception can be computed as (\ref{gamma}), which yields (\ref{SINR}) after some simple manipulations.

Therefore, by letting $\bar{n}=2N_{\textrm{T}}$, $t=N_{\textrm{T}}$, $\bar{\textbf{C}}_{\textbf{G}}=\bar{\bar{\textbf{X}}}_{i+1,\textbf{G}}$, $\bar{\textbf{C}}_{{\textbf{G}}'}=\bar{\bar{\textbf{X}}}_{i+1,{\textbf{G}}'}$, $r=N_{\textrm{R}}-l$, and $\rho=\gamma$ in (\ref{Original-PEP}), we complete the proof.

\setcounter{equation}{48}

\subsection{Proof of Theorem 3}\label{proof_theorem3}
Define $\textbf{I}_{N_{\textrm{R}},l}$ as the $N_{\textrm{R}} \times N_{\textrm{R}}$ matrix with $1$ on its $l$-th superdiagonal and 0 in other entries. On this basis, when the correlation matrix $\textbf{C}$ has only one superdiagonal (or main diagonal), we can recast $\textbf{C}$ as $\rho_{l}\textbf{I}_{N_{\textrm{R}},l}$. Therefore, (\ref{Improved-Decoding}) can be simplified as
\begin{equation}
\begin{split}
\widehat{\textbf{G}}
=&\mathrm{arg}\underset{\textbf{G}\in \mathcal{G}}{\mathrm{max}}\mathrm{Re}\left\{ \mathrm{Tr} \left\{\textbf{G}\textbf{Y}_{i+1}^{H}\left( \rho_{l}\textbf{I}_{N_{\textrm{R}},l} \right) \textbf{Y}_{i} \right\}\right\}\\
=&\mathrm{arg}\underset{\textbf{G}\in \mathcal{G}}{\mathrm{max}}\mathrm{Re}\left\{ \mathrm{Tr} \left\{\textbf{G}\textbf{Y}_{i+1}^{H}\textbf{I}_{N_{\textrm{R}},l}\textbf{Y}_{i} \right\}\right\}\\
=&\mathrm{arg}\underset{\textbf{G}\in \mathcal{G}}{\mathrm{max}}\mathrm{Re}\left\{ \mathrm{Tr} \left\{\textbf{G}\textbf{Y}_{i+1}^{H}\left[ \begin{matrix}
\left[ \textbf{Y}_{i} \right ]_{\left(l+1 \right ):N_{\textrm{R}}}
\\
\textbf{0}
\end{matrix} \right] \right\}\right\}\\
=&\mathrm{arg}\underset{\textbf{G}\in \mathcal{G}}{\mathrm{max}}\mathrm{Re}\left\{ \mathrm{Tr} \left\{\textbf{G}\left( \left[ \textbf{Y}_{i+1} \right ]_{1:\left(N_{\textrm{R}}-l \right )}\right )^{H}\left[ \textbf{Y}_{i} \right ]_{\left(l+1 \right ):N_{\textrm{R}}} \right\}\right\},
\label{Improved-Decoding-pro}
\end{split}
\end{equation}
where $\left[ \textbf{A} \right]_{m:n}$ denotes the matrix that contains the $m$ to $n$-th rows of $\textbf{A}$.

Next, define $\bar{\textbf{Y}}_{i}\triangleq \left[ \textbf{Y}_{i} \right ]_{\left(l+1 \right ):N_{\textrm{R}}}$ and $\bar{\textbf{Y}}_{i+1}\triangleq \left[ \textbf{Y}_{i+1} \right ]_{1:\left(N_{\textrm{R}}-l \right )}$, we have
\begin{equation}
\widehat{\textbf{G}}=\mathrm{arg}\underset{\textbf{G}\in \mathcal{G}}{\mathrm{max}}\mathrm{Re}
\left\{
\mathrm{Tr}
\left\{
\textbf{G} \bar{\textbf{Y}}_{i+1}^{H} \bar{\textbf{Y}}_{i}
\right\}
\right\}.
\label{Standard-Decoding3}
\end{equation}
Notice that $\bar{\textbf{Y}}_{i}$ corresponds to the signal received by antennas in set $\left\{ l+1,\dots,N_{\textrm{R}} \right\}$ at time $t_{i}$, and $\bar{\textbf{Y}}_{i+1}$ represents the signal received by antennas $\left\{ 1,\dots,N_{\textrm{R}}-l \right\}$ at time $t_{i+1}$, (\ref{Standard-Decoding3}) is equivalent to (\ref{Standard-Decoding}) with $N_{\textrm{R}}-l$ receive antennas.

\subsection{Proof of Theorem 4}\label{proof_theorem4}
\newcounter{TempEqCnt5}
\setcounter{TempEqCnt5}{\value{equation}}
\setcounter{equation}{54}
\begin{figure*}[ht]
\begin{equation}
\begin{split}
&\textrm{Pr}\left\{ \textbf{G}\rightarrow \textbf{G}' \right\}=\textrm{Pr}\left\{ \mathrm{Re}\left \{ \mathrm{Tr} \left\{\left(\textbf{G}'- \textbf{G} \right)\bar{\textbf{Y}}_{i+1}^{H}\bar{\textbf{Y}}_{i} \right\}\right\} >0 \right\}\\
&=\textrm{Pr}\left\{ \mathrm{Re}\left \{ \mathrm{Tr} \left\{\left(\textbf{G}'- \textbf{G} \right)\left( \sqrt{P}\bar{\textbf{H}}_{i}\textbf{X}_{i}\textbf{G}+\sqrt{\frac{N_{\textrm{T}}}{\gamma_2}}\tilde{\textbf{N}}_{2} \right)^{H}\left( \sqrt{P} \bar{\textbf{H}}_{i}\textbf{X}_{i}+\sqrt{\frac{N_{\textrm{T}}}{\gamma_1}}\tilde{\textbf{N}}_{1} \right) \right\}\right\} >0 \right\}\\
&= \textrm{Pr}\left\{ \mathrm{Re}\left \{ \mathrm{Tr} \left\{\left(\textbf{G}'- \textbf{G} \right)\left( P\textbf{G}^{H}\textbf{X}_{i}^{H}\bar{\textbf{H}}_{i}^{H}\bar{\textbf{H}}_{i}\textbf{X}_{i}+\sqrt{\frac{N_{\textrm{T}}P}{\gamma_2}}\tilde{\textbf{N}}_{2}^{H}\bar{\textbf{H}}_{i}\textbf{X}_{i} + \sqrt{\frac{N_{\textrm{T}}P}{\gamma_1}}\textbf{G}^{H}\textbf{X}_{i}^{H}\bar{\textbf{H}}_{i}^{H}\tilde{\textbf{N}}_{1} + \frac{N_{\textrm{T}}}{\sqrt{\gamma_{1} \gamma_{2}}}\tilde{\textbf{N}}_{2}^{H}\tilde{\textbf{N}}_{1}  \right) \right\}\right\} >0 \right\}\\
&\leq  \textrm{Pr}\left\{ \mathrm{Re}\left \{ \mathrm{Tr} \left\{\left(\textbf{G}'- \textbf{G} \right)\left( P \textbf{G}^{H}\textbf{X}_{i}^{H}\bar{\textbf{H}}_{i}^{H}\bar{\textbf{H}}_{i}\textbf{X}_{i}+\sqrt{\frac{N_{\textrm{T}}P}{\gamma}}\tilde{\textbf{N}}_{2}^{H}\bar{\textbf{H}}_{i}\textbf{X}_{i} + \sqrt{\frac{N_{\textrm{T}}P}{\gamma}}\textbf{G}^{H}\textbf{X}_{i}^{H}\bar{\textbf{H}}_{i}^{H}\tilde{\textbf{N}}_{1} + \frac{N_{\textrm{T}}}{\gamma}\tilde{\textbf{N}}_{2}^{H}\tilde{\textbf{N}}_{1}  \right) \right\}\right\} >0 \right\}\\
&= \textrm{Pr}\left\{ \mathrm{Re}\left \{ \mathrm{Tr} \left\{\left(\textbf{G}'- \textbf{G} \right)\left( \sqrt{P} \bar{\textbf{H}}_{i}\textbf{X}_{i}\textbf{G}+\sqrt{\frac{N_{\textrm{T}}}{\gamma}}\tilde{\textbf{N}}_{2} \right)^{H}\left( \sqrt{P} \bar{\textbf{H}}_{i}\textbf{X}_{i}+\sqrt{\frac{N_{\textrm{T}}}{\gamma}}\tilde{\textbf{N}}_{1} \right) \right\}\right\} >0 \right\}\\
&=\textrm{Pr}\left\{ \mathrm{Re}\left \{ \mathrm{Tr} \left\{\left(\textbf{G}'- \textbf{G} \right)\tilde{\textbf{Y}}_{i+1}^{H}\tilde{\textbf{Y}}_{i} \right\}\right\} >0 \right\}
\end{split}\label{PEP-expression}
\end{equation}
\hrulefill
\end{figure*}
\setcounter{equation}{\value{TempEqCnt5}}

Invoking the results in Appendix \ref{proof_theorem3}, we see only some particular rows of $\textbf{Y}_{i}$ and $\textbf{Y}_{i+1}$, i.e., $\bar{\textbf{Y}}_{i}=\left[ \textbf{Y}_{i} \right ]_{\left(l+1 \right ):N_{\textrm{R}}}$ and $\bar{\textbf{Y}}_{i+1}=\left[ \textbf{Y}_{i+1} \right ]_{1:\left(N_{\textrm{R}}-l \right )}$, are involved in differential decoding. For further analysis, we define their corresponding channel matrices as $\bar{\textbf{H}}_{i}\triangleq \left[ \textbf{H}_{i} \right ]_{\left(l+1 \right ):N_{\textrm{R}}}$ and $\bar{\textbf{H}}_{i+1}\triangleq \left[ \textbf{H}_{i+1} \right ]_{1:\left(N_{\textrm{R}}-l \right )}$, respectively. Recalling that $\textbf{C}=\rho_{l}\textbf{I}_{N_{\textrm{R}},l}$ and $\textbf{H}_{i+1}=\textbf{C}\textbf{H}_{i}+\textbf{U}_{i+1}$, it follows that
\begin{equation}
\bar{\textbf{H}}_{i+1}=\rho_{l}\bar{\textbf{H}}_{i}+\bar{\textbf{U}}_{i},
\label{Overall-AR1-Model-pro}
\end{equation}
where $\bar{\textbf{U}}_{i}\triangleq\left[ \textbf{U}_{i} \right ]_{\left(l+1 \right ):N_{\textrm{R}}}$ and each column of $\bar{\textbf{U}}_{i}$ has the covariance matrix $\textbf{R}_{\bar{\textrm{u}}}=\left( 1-\rho_{l}^{2} \right) \textbf{I}_{N_{\textrm{R}}-l}$. Based on these definitions, (\ref{Two-Received}) can be rewritten as
\begin{equation}
\begin{split}
\bar{\textbf{Y}}_{i}~=~&\sqrt{P}\bar{\textbf{H}}_{i}\textbf{X}_{i}+\bar{\textbf{N}}_{i},\\
\frac{1}{\rho_{l}}\bar{\textbf{Y}}_{i+1}~=~&\sqrt{P}\bar{\textbf{H}}_{i}\textbf{X}_{i}\textbf{G}_{k}+\frac{\sqrt{P}}{\rho_{l}}\bar{\textbf{U}}_{i}\textbf{X}_{i}\textbf{G}_{k}+\frac{1}{\rho_{l}}\bar{\textbf{N}}_{i+1},
\label{Received-Symbol-pro}
\end{split}
\end{equation}
where $\bar{\textbf{N}}_{i} \triangleq \left[ \textbf{N}_{i} \right ]_{\left(l+1 \right ):N_{\textrm{R}}}$ and $\bar{\textbf{N}}_{i+1} \triangleq \left[ \textbf{N}_{i+1} \right ]_{1:\left( N_{\textrm{R}}-l \right)}$. Hence, the SINRs associated with $\bar{\textbf{Y}}_{i}$ and $\bar{\textbf{Y}}_{i+1}$, denoted as $\gamma_{1}$ and $\gamma_{2}$ respectively, can be easily derived as
\begin{equation}
\gamma_{1}=\frac{N_{\textrm{T}}P}{\sigma^{2}_{\emph{n}}},~~~~~
\gamma_{2}=\frac{\left| \rho_{l} \right|^{2}}{1-\left| \rho_{l} \right|^{2}+\frac{\sigma^{2}_{\emph{n}}}{N_{\emph{T}}P}}.
\label{SINR-pro-backup}
\end{equation}
Through normalization, the interference-plus-noise counterparts in (\ref{Received-Symbol-pro}) can be presented as
\begin{equation}
\sqrt{\frac{N_{\textrm{T}}}{\gamma_1}}\tilde{\textbf{N}}_{1}=\bar{\textbf{N}}_{i}, ~~
\sqrt{\frac{N_{\textrm{T}}}{\gamma_2}}\tilde{\textbf{N}}_{2}=\frac{\sqrt{P}}{\rho_{l}}\bar{\textbf{U}}_{i}\textbf{X}_{i}\textbf{G}_{k}+\frac{1}{\rho_{l}}\bar{\textbf{N}}_{i+1},
\label{IN-part}
\end{equation}
where the entries of $\tilde{\textbf{N}}_{1}$ and $\tilde{\textbf{N}}_{2}$ are i.i.d. complex Gaussian variables with zero-mean and unit variance. Substituting (\ref{Received-Symbol-pro}) and (\ref{IN-part}) into (\ref{Standard-Decoding3}), the PEP can be expressed as in (\ref{PEP-expression}), where
$\tilde{\textbf{Y}}_{i}\triangleq\sqrt{P} \bar{\textbf{H}}_{i}\textbf{X}_{i}+\sqrt{\frac{N_{\textrm{T}}}{\gamma}}\tilde{\textbf{N}}_{1}$, $\tilde{\textbf{Y}}_{i+1}\triangleq\sqrt{P} \bar{\textbf{H}}_{i}\textbf{X}_{i}\textbf{G}+\sqrt{\frac{N_{\textrm{T}}}{\gamma}}\tilde{\textbf{N}}_{2}$,
and $\gamma=\frac{2\gamma_{1}\gamma_{2}}{\gamma_{1}+\gamma_{2}}$ represents the corresponding equivalent SINR. Note that the inequality in (\ref{PEP-expression}) holds due to the fact that, the error probability is mainly determined by the last three independent noise terms within the parentheses, and replacing $\gamma_{1}$ and $\gamma_{2}$ with $\gamma$ results in a larger sum noise power (the sum of the first two remains the same but the third one becomes larger) which means a larger average error probability.

Consequently, we can get the PEP upper bound by simply executing the PEP of the equivalent unitary coding system defined by $\tilde{\textbf{Y}}_{i}$ and $\tilde{\textbf{Y}}_{i+1}$ with an SNR of $\gamma$, which just yields (\ref{PEP-pro}).

\section*{Acknowledgements}
The authors sincerely thank the Editor and the anonymous reviewers for their constructive suggestions which helped us to improve the manuscript.

\bibliographystyle{IEEEtran}

\end{document}